\documentclass[referee]{raa}

\usepackage{graphicx,times}           
\usepackage{threeparttable}
\usepackage{natbib}
\usepackage{amssymb,amsmath}
  \usepackage{latexsym}
  \usepackage{longtable}
  \usepackage{verbatim}
  \usepackage{mathtools}
  \usepackage{multirow}
\usepackage{empheq}
\usepackage{bm}
\usepackage{color}
\usepackage{CJK}
\usepackage{longtable}

\bibpunct{(}{)}{;}{a}{}{,}
\usepackage{hyperref}
\hypersetup{pdftitle = The title of my PDF, pdfauthor = My name, pdfsubject= The subject, pdfkeywords = keyword1 keyword2 keyword3} 
\hypersetup{colorlinks = true, linkcolor = green, anchorcolor = red, citecolor = blue, filecolor = red, pagecolor = red, urlcolor = red}

\newcommand{\Ha}{H$\alpha$\,$\lambda$6563}
\newcommand{\Hb}{H$\beta$\,$\lambda$4861}

\newcommand{\OIII}{[O\,{\sc iii}]\,$\lambda$5007}

\newcommand{\OIIII}{[O\,{\sc iii}]}
\newcommand{\LOIII}{$L_{\rm [O\,{\scriptsize \textsc{iii}}]}$}
\newcommand{\NII}{[N\,{\sc ii}]\,$\lambda$6584}
\newcommand{\NIII}{[N\,{\sc ii}]}

\begin{document}
\title{Radial stellar populations of  AGN-host dwarf galaxies in SDSS-IV MaNGA survey}

  \volnopage{Vol.0 (20xx) No.0, 000--000}      
   \setcounter{page}{1}          

   \author{Wei Cai
      \inst{1,2,3}
   \and Yinghe Zhao
      \inst{1,3,4}\footnote{Author to whom any correspondence should be addressed}
    \and Jin-Ming Bai
      \inst{1,3,4}  
   }

   \institute{Yunnan Observatories, Chinese Academy of Sciences, Kunming 652016, China; {\it zhaoyinghe@ynao.ac.cn}\\
        \and
            University of Chinese Academy of Sciences, Beijing 100049, China\\
        \and
           Key Laboratory for the Structure and Evolution of Celestial Objects, Chinese Academy of Sciences, Kunming 650216, China\\
         \and
            Center for Astronomical Mega-Science, Chinese Academy of Sciences, Beijing 100012, China\\
\vs\no
   {\small Received~~20xx month day; accepted~~20xx~~month day}}

\abstract{
Based on  MaNGA integral field unit (IFU) spectroscopy  we  search 60 AGN candidates, which have stellar masses $M_{\star}\leqslant5\times10^{9}$$M_{\odot}$  and show AGN ionization signatures in the BPT diagram.  For these AGN candidates, we derive the spatially resolved stellar population with the stellar population synthesis code STARLIGHT and  measure the gradients of the mean  stellar age and metallicity. We find that the gradients of mean stellar age (metallicity) of individual AGN-host dwarfs are diverse  in 0$-$0.5 $Re$, 0.5$-$1 $Re$ and 0$-$1 $Re$. However, the overall behavior of the mean stellar age  (metallicity) profiles tend to be flat, as the median values of the gradients are close to zero. We further study  the overall behavior of the mean stellar age (metallicity) by plotting the co-added radial profiles for the AGN sample and compare with a control sample  with similar stellar mass. We find that the median values of light-weighted mean stellar ages of AGN sample  are as old as 2$-$3 ~Gyr within 2 $Re$,which are about 4$-$7 times older than those of the control sample. Meanwhile, most of the AGN candidates are low-level AGNs, as only eight sources have $L_{\rm [O\,{\scriptsize \textsc{iii}}]}$$>$$10^{39.5}$~erg~s$^{-1}$. Hence, the AGNs in dwarf galaxies might  accelerate the evolution of galaxies by accelerating the consumption of the gas, resulting in an overall quenching of the dwarf galaxies, and the AGNs also become weak due to the lack of gas. The median values of mass-weighted mean stellar age of both samples within 2 $Re$ are similar and as old as  about 10~Gyr, indicating  that the stellar mass is mainly contributed by old stellar populations.   The gradients of co-added mean stellar metallicity for both samples tend to be negative but close to zero, and the similar mean stellar metallicity profiles for both samples  indicate that the chemical evolution of the host galaxy is not strongly influenced by the AGN.
\keywords{galaxies: dwarf - galaxies: active - galaxies: galaxies stellar content}
}

   \authorrunning{Wei Cai et al.}            
   \titlerunning{Radial stellar populations of  AGN-host dwarf galaxies in SDSS-IV MaNGA survey }  

   \maketitle

\section{Introduction}
 \label{sec:intro}
The relation between the mass of supermassive black holes (SMBHs) and bulge stellar velocity dispersion (the $M_\star-\sigma$ relation) suggests that the growth of central SMBHs is coupled with the growth of host galaxies (e.g., \citealt{Ferrarese+Merritt+2000, Gebhardt+etal+2000, Kormendy+Ho+2013}).  This relation has been explained by  active galactic nuclei (AGN) feedback process that  AGNs regulate the star formation of host galaxies  in both radiative and kinetic forms (e.g., \citealt{Fabian+2012}), which might heat the gas, or even expel the gas from the galaxy. 
Indeed, 
the feedback mechanisms such as AGNs (e.g., \citealt{Bower+etal+2006, Schaye+etal+2015, Terrazas+etal+2020}) must be included in cosmological models to regulate star formation of the massive galaxies. Therefore, AGN effects on the evolution of massive galaxies have been widely accepted (see the reviews by \citealt{Heckman+Best+2014}). 

AGNs have also been found in low-mass, dwarf galaxies (e.g., NGC 4395; \citealt{Filippenko+Sargent+1989}; POX 52; \citealt{Barth+etal+2004}).  In recent years, more and more dwarf galaxies with clear AGN signatures are identified  at all wavelength from X-ray to radio band (e.g., \citealt{Reines+etal+2013, Moran+etal+2014, Lemons+etal+2015, Sartori+etal+2015, Mezcua+etal+2019, Birchall+etal+2020, Reines+etal+2020}).  The central BHs in dwarf galaxies are suggested to be intermediate-mass black holes (IMBHs)  with mass between $10^2$ and $10^6 M_\odot$ (e.g., \citealt{Reines+etal+2013, Moran+etal+2014, Silk+2017}), which might have similar mass with the first seed black holes due to the relatively quiet merger histories of dwarf galaxies (\citealt{Bellovary+etal+2011}).  Furthermore, dwarf galaxies have similar conditions of high-redshift galaxies due to their low metallicity (e.g., \citealt{Mateo+1998}).  Therefore, it's essential to investigate  the AGN effects on the dwarf galaxies, which could  improve our understanding of AGN role in galaxy formation and evolution.

In recent years,  a series of works have studied the AGN effects on the evolution of dwarf galaxies, while the importance of AGNs in regulating the star formation of dwarf galaxies is still in debate.  
  Some works suggest that AGNs should have strong effects on the star formation of dwarf galaxies (e.g., \citealt{Dashyan+etal+2018, Penny+etal+2018, Barai+deGouveia+2019, Manzanoking+etal+2019, Mezcua+etal+2019, Sharma+etal+2020}). However, other works (e.g., \citealt{Trebitsch+etal+2018, Koudmani+etal+2019}) propose that AGN effects on regulating the star formation of host galaxies are negligible. 

AGN is expected to have connection with the stellar population of  host galaxy.  AGN may be triggered when the nuclear engine is  being fed by the inflow of material, which could also be  used to form new stars. AGN feedback might also influence the stellar population of host galaxy by regulating the star formation.    A number of works have studied the stellar populations of AGN-host massive galaxies by using both central spectra (e.g., \citealt{CidFernandes+etal+2004}; hereafter C04) and IFU spectra (e.g., \citealt{Sanchez+etal+2018, Mallmann+etal+2018}) and suggested that nuclear activities might be relevant to recent episodes of circumnuclear star formation (e.g., \citealt{StorchiBergmann+etal+2001, Rembold+etal+2017}).   In AGN-host dwarfs, our  work (\citealt{Cai+etal+2020}) presents a detailed study of the stellar populations for a sample containing 136 sources and shows a diversity of SFHs for these optically selected AGN-host dwarfs.  We further find a  mild correlation between the SFHs and the luminosity of the  \OIII\ line ($L_{\rm [O\,{\scriptsize \textsc{iii}}]}$) for sources with $L_{\rm [O\,{\scriptsize \textsc{iii}}]} > 10^{39}$~erg~s$^{-1}$,  indicating a physical connection between star formation and AGN activities.

 However, the stellar populations of the AGN-host dwarfs in \cite{Cai+etal+2020} are limited to the central regions using the spectra observed from a 3\arcsec-diameter aperture. In recent years,  the IFU technology has been developed and  the integral-field spectroscopy can provide us  a wealth of information for each individual galaxy (such as the  spatially resolved stellar populations, which  can be used to locate the sites of new and recent star formation), which can improve our understanding of the evolution of galaxies and reveal the possible quenching mechanisms. The  spatial distribution of the stellar ages, element abundances, and star formation histories (or the gradients of these parameters) can reveal how galaxy grows or quenchs.   Many works have studied the  age and metallicity gradients for both disc and elliptical galaxies, and most of their works (e.g., \citealt{GonzalezDelgado+etal+2015, Morelli+etal+2015, Zheng+etal+2017}) find a negative gradient,  indicating inside-out growth of galaxies.   However,  the outside-in mode has been suggested in many nearby dwarf galaxies (e.g., \citealt{zhang+etal+2012}), and a negative correlation is found between the  stellar population gradients  and the stellar mass (e.g., \citealt{ Chen+ertal+2020}). A series of works (e.g., \citealt{ Zheng+etal+2017, Chen+ertal+2020}) have also found no/weak correlation between the age (metallicity) gradients and the local density environment .


To further investigate the AGN effects on the SFHs of dwarf galaxies, in this paper we present our detailed studies on the radial stellar populations for a sample of 60 AGN-host dwarf galaxies, selected from the MaNGA survey. We further compare the results of AGN-host dwarfs with a control sample of carefully selected normal galaxies. The paper is organized as follows. Section 2 describes  the basic informations of MaNGA data, the selection of the AGN  and control samples and our data reductions. Our results and analysis are given in Section 3.  We compare the properties of the AGN sample with those of control samples, the H$\alpha$ equivalent width of the AGN sample in Section 4 and present  our conclusions in the last section. Throughout this paper, we adopt a Hubble constant of $H_{0}$ = 70 km s$^{-1}$Mpc$^{-1}$,
$\Omega_{M}$=0.28, and $\Omega_{\Lambda}$=0.72, which are based on the five-year \textit{WMAP} results (\citealt{Hinshaw+etal+2009}).

\section{SAMPLE AND DATA ANALYSIS}
\subsection{Sample}

MaNGA is the largest ongoing IFU survey, aiming to acquire spatially resolved spectra of  100 00 nearby galaxies ($z <$ 0.1) from 2014 to 2020 (\citealt{Bundy+etal+2015, Law+etal+2015, Blanton+etal+2017}). MaNGA target galaxies are selected from the NASA-Sloan Atlas (NSA) catalog \footnote{https://www.sdss.org/dr15/manga/{manga-target-selection}/nsa/ } (version v1\_0\_1) with a flat distribution in the stellar mass above 10$^9 M_{\odot}$.  In addition, a small ancillary sample is observed to extend the MaNGA sample to lower mass. 
The  MaNGA observations are carried out with 17 fiber-bundle IFUs  with 5 different sizes, varying in diameter from 12\arcsec (19 fibers) to 32\arcsec (127 fibers). 
Galaxies are covered to at least 1.5 effective radii ($Re$) in observation, where $Re$ represents the radius containing 50\% of the light of the galaxy measured at r-band.
The observational spectrum has a  spectral coverage ranging from 3600\r{A} to 10300 \r{A} with a spectral resolution varying from R $\sim$ 1400 at 4000\r{A} to R $\sim$ 2600 around 9000\r{A}  (\citealt{Drory+etal+2015, Yan+etal+2016a, Yan+etal+2016b}) provided by the dual beam BOSS spectrographs (\citealt{Smee+etal+2013}).

The MaNGA parent sample used here  is from the internal MaNGA Product Launch 6 (MPL 6) with 4718 galaxies observed within the first 4 years, corresponding to the SDSS data release 15\footnote{http://www.sdss.org/surveys/manga}. The raw data were reduced, calibrated, and reconstructed to a data cube by the Data Reduction Pipeline (DRP) (\citealt{Law+etal+2016}). The final 3D data cubes deliver the spectra for each spaxel of a given galaxy and each spaxel has an area of 0\arcsec.5$\times$0\arcsec.5. 

\subsubsection{AGN Sample}

\begin{figure}[t!]
\centering
\includegraphics[ width=0.95\textwidth]{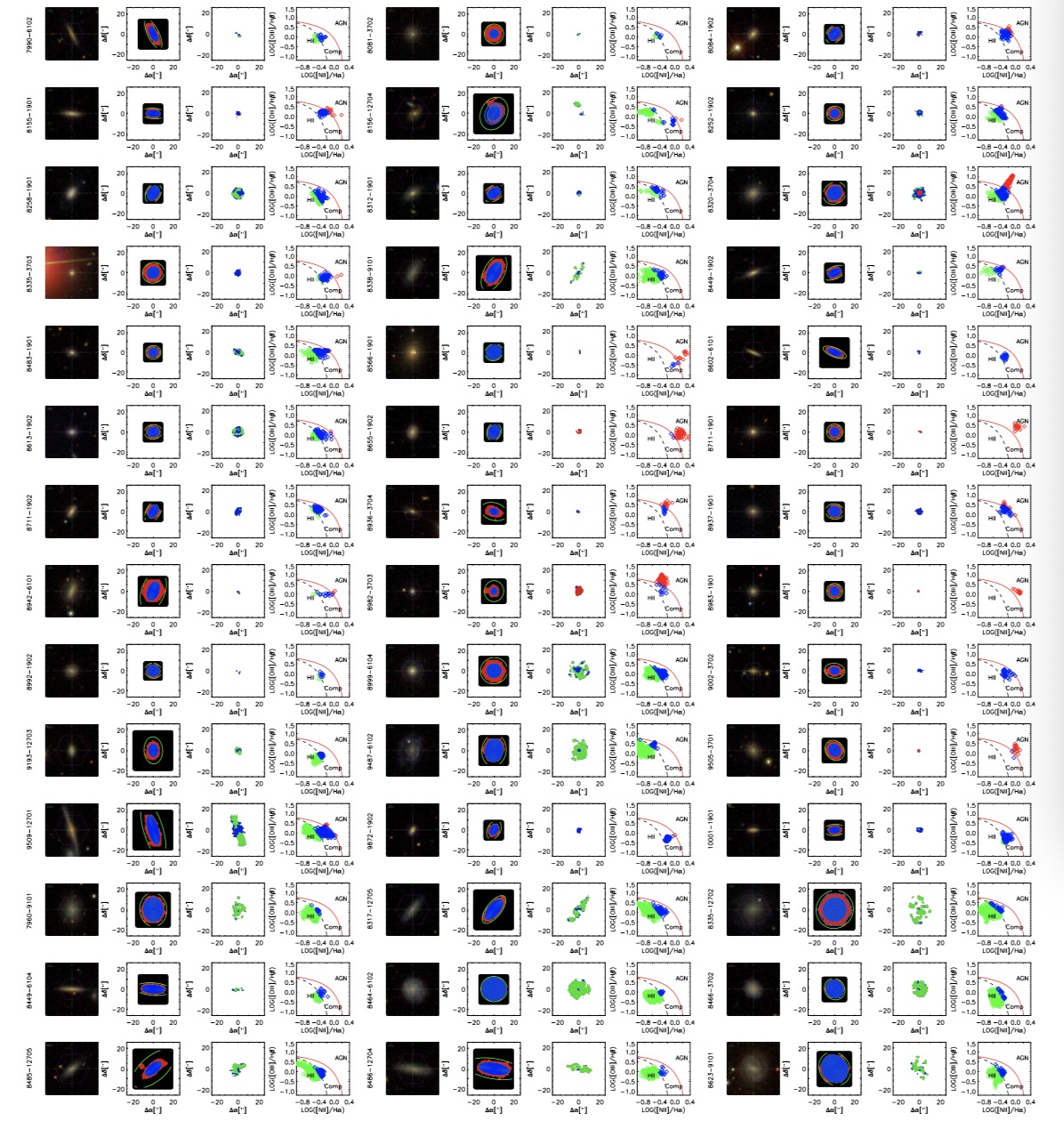}
\caption{The 49 AGN candidates identified by the spatially resolved BPT diagram.  From left to right: the gri false-color images of AGN-host galaxies with MaNGA IFU footprint overlaid; The contour with the elliptical annuli we adopted to increase the S/N by binning the spectra. The blue elliptical annuli represent that the S/N is larger than 10, whereas the green elliptical annuli represent that the S/N is lower than 10 for the external regions of most galaxies; The BPT maps of AGN-host galaxies based on the emission line flux maps produced by MaNGA DAP; The BPT  diagrams of AGN-host galaxies with different colors (AGNs (red), composite (blue) and star formation (green)) representing  the ionization in different part of galaxies. The continuous lines separating AGNs, composite objects and star-forming galaxies are from \cite{Kauffmann+etal+2003} and \cite{Kewley+etal+2001}. }
\label{figage}
 \end{figure}

  \addtocounter{figure}{-1}

\begin{figure}[t!]
\centering
\includegraphics[ width=0.95\textwidth]{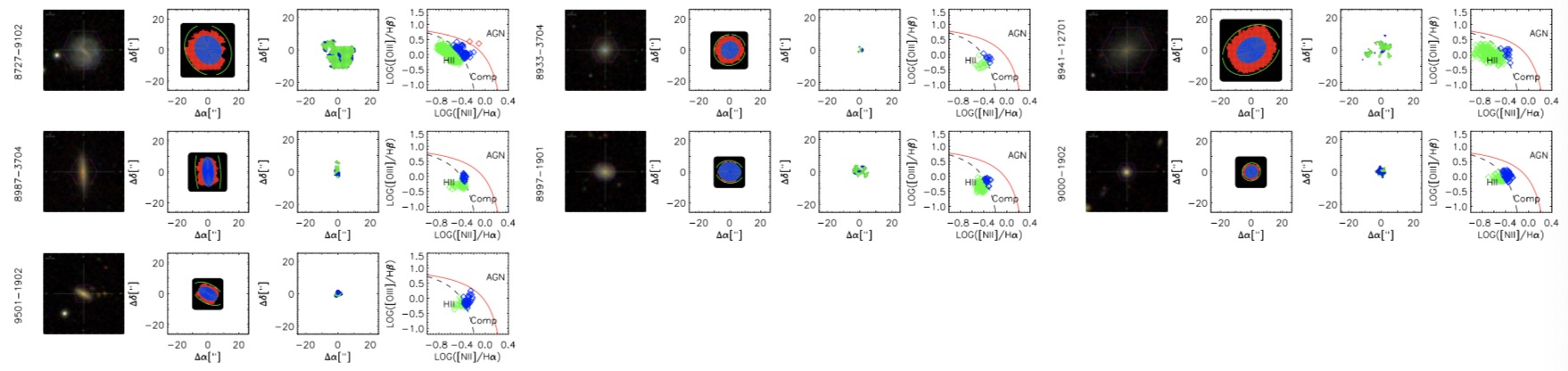}
\caption{Continued }
\label{figage}
 \end{figure}

\begin{figure}[t!]
\centering
\includegraphics[width=0.45\textwidth]{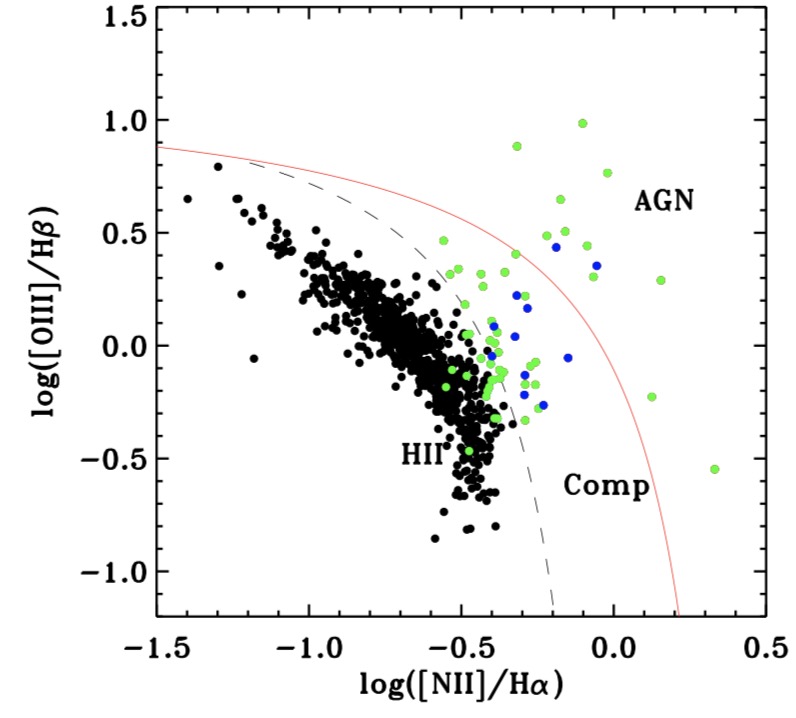}
\caption{BPT diagram for the dwarf galaxies using the emission lines fluxes from the pure central 3\arcsec\   spectra. The continuous lines separating AGNs, composite objects and star-forming galaxies are  from \cite{Kauffmann+etal+2003} and \cite{Kewley+etal+2001}. Colored dots represent the 60 AGN candidates.  Green dots represent the 49 AGN candidates identified by the spatially resolved BPT diagram. }
\label{bptfigure}
 \end{figure}

\begin{figure*}[t!]
\centering
\includegraphics[ width=0.85\textwidth]{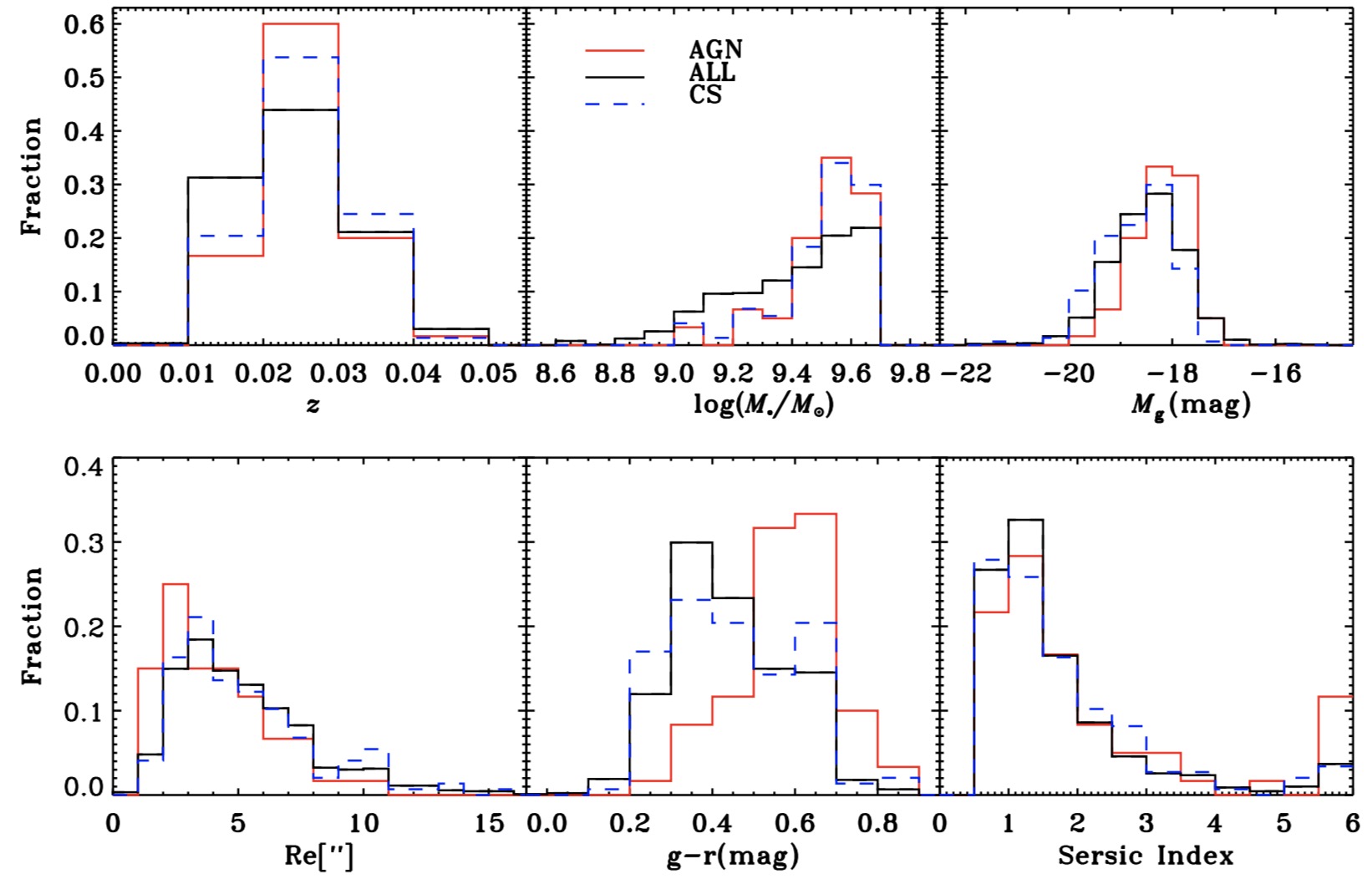}
\caption{The distributions of redshifts, stellar mass, absolute $g$ band magnitude,  effective radius, the $g-r$ color and S\'{e}rsic index of the AGN sample (red), the control sample (blue) and the all dwarf galaxies (black).  The $g-r$ color of AGN sample is redder than that of the whole dwarf galaxies. All values are from the NSA catalog v1\_0\_1 and assume h = 0.7. Magnitudes are K-corrected to rest-frame values and corrected for foreground Galactic extinction. }
\label{comparefigure}
\end{figure*}

To select dwarf galaxies, we follow similar stellar mass limits as other work in the field (e.g.,  \citealt{Reines+etal+2013}; \citealt{Penny+etal+2016,Penny+etal+2018}) , i.e., $M_{\star} \leq$ 5$\times$10$^{9} M_{\odot}$, corresponding roughly to the stellar mass of the Large Magellan Cloud. 
The stellar mass adopted here is derived with elliptical Petrosian fluxes (NSA\_ELPETRO\_MASS) from the MaNGA parent sample.  There are about 900 dwarf galaxies in the MaNGA parent sample. We identify  AGN candidates from this dwarf sample using the Baldwin, Phillips, \& Telervich (BPT) diagram (e.g., \citealt{Baldwin+etal+1981, Veilleux+Osterbrock+1987}), which  adopt the 2D line-intensity ratio calculated from relatively strong lines of \Hb, \OIII, \Ha,  and \NII\  to probe the nebular conditions of a source.  \cite{Mezcua+etal+2020} have suggested that  the AGN emission could be off-nuclear and the AGN emission  might be overwhelmed by the light from its host galaxy when  the  integrated spectra are used. Therefore, for each dwarf galaxy, we construct the spatially resolved BPT diagram.   Spaxels with  signal-to-noise ratios (S/N) of the above four emission lines larger than 3  where are measured with the non-parametric emission line  by MaNGA DAP,  are used. A galaxy is considered as an AGN when the circumnuclear  regions are classified as either the composite or the AGN.  Finally,  49  AGN candidates are found by the spatially resolved BPT diagram, as shown in Figure \ref{figage}.

 Since we require all of the four lines having their S/N $>$ 3, it might miss some AGNs using the spatially resolved BPT diagram. 
 Therefore we bin the central 1\arcsec.5 radius spectra for each galaxy and   model the stellar continuum using the STARLIGHT code (see \S2.3).  The pure emission line spectrum is obtained by subtracting the best-fit model spectrum from the observed one.  We fit the emission lines with a Gaussian function, get the fluxes of  H$\alpha$, H$\beta$, \OIIII, \NIII, and make use of the BPT diagram to classify the galaxy. Finally,  44 AGN-host galaxies (13 have AGN dominated spectra and 31 have composite spectra) are identified, as shown in Figure \ref{bptfigure}.  The S/N of the four emission lines are all greater than 3 for these sources.  There are 33 sources are overlapped between the central-3\arcsec and spatially-resolved samples and thus the final sample contains 60 sources. We further checked the 16 AGNs which are classified as SF galaxies by using the central-3\arcsec spectrum and found that most spaxels with AGN signature are off-nuclear.   The existence of off-center AGNs  are consistent with  \cite{Bellovary+2019} results, who have predicted that  the wandering  black holes are presented in dwarf galaxies, which might be attributed to the physical effects, such as interactions and mergers.  However, these off-center AGNs might be not true off-center AGNs due to the star formation dilution in the center of the galaxy. It needs more data, such as X-ray , to further investigate the nature of these  off-nuclear AGNs.

   In table \ref{agnsm}, we show the properties for these AGN candidates.  Four out of six AGN dwarf galaxies found by \cite{Penny+etal+2018} are included in our final AGN sample. One source (7815$-$1901) is not included because its mass is larger than  $5\times10^9 M_\odot$ and the another one (8623$-$9102) is caused by the different criterion in selecting the spaxels to construct the spatially resolved BPT diagram. Our AGN sample has a median redshift of $\langle z \rangle = 0.025$, a median $g$-band absolute magnitude of $\langle M_g \rangle \sim -18.2$ mag, a median $g-r$ color of $\langle g-r \rangle =0.58$ mag, and a median stellar mass of $\langle M_\star \rangle =3.5\times10^9\,M_\odot$.  These properties of our AGN sample are similar to the larger  AGN host dwarf galaxy sample in \cite{Reines+etal+2013}. Meanwhile, as shown in  Figure \ref{comparefigure},  we find that the distributions of  galactic parameters of our AGN sample are similar to those of the parent sample, except for that  few AGNs  are found at the blue end.   \cite{Reines+etal+2013} also found that their AGN sample selected by BPT diagram tends to be redder compared to the parent sample of dwarfs (see their Figure 7 for details). \cite{Trump+etal+2015} have suggested that AGN signature might be overwhelmed by the light from its host galaxy. Therefore, the redder color of AGN sample might be due to a selection bias arising from the optical diagnostics, which is not sensitive to AGNs in blue star-forming low-mass dwarf galaxies  since the observed emission is dominated by galaxy light.
 
 \begin{longtable}{ccccccccccc}
\caption{Properties of the 60 AGN-host dwarf galaxies.} \label{agnsm}\\
\hline

manga-Plateifu  & manga-ID & Ra& Dec & $z$ &  log $M_{\star}$& $M_g$&$Re$ & g-r &  S\'ersic  n &EW($H\alpha$) \\
 & & (J2000)&(J2000)& &$M_\odot$&(mag)&(kpc)& & &(\r{A})\\
(1)&(2)&(3)&(4)&(5)&(6)&(7)&(8)&(9)&(10)&(11)\\
\endfirsthead
\hline
manga-Plateifu  & manga-ID & Ra& Dec & $z$ &  log $M_{\star}$& $M_g$&$Re$ & g-r &  S\'ersic  n &EW($H\alpha$) \\
 & & (J2000)&(J2000)& &$M_\odot$&(mag)&(kpc)& & &(\r{A})\\
(1)&(2)&(3)&(4)&(5)&(6)&(7)&(8)&(9)&(10)&(11)\\ \endhead
\hline
\endfoot

\endlastfoot
\hline
 \multicolumn{11}{c}{The 33 AGN candidates using both spatially resolved and central 1\arcsec.5 radius spectra }\\
 \hline
 7990$-$6102    &   1$-$24379    &  261.466028    &   56.852051    &     0.02679    &        9.44    &      -17.68    &        3.10    &        0.66    &         1.1    &        5.69   \\
 8081$-$3702$^{c}$    &   1$-$38166    &   49.946854    &    0.623822    &     0.02465    &        9.45    &      -18.58    &        2.23    &        0.37    &         3.4    &        2.64   \\
$^a$ 8084$-$1902$^{c}$    &   1$-$38618    &   52.622585    &   -0.488769    &     0.02173    &        9.67    &      -18.54    &        1.02    &        0.57    &         2.7    &        2.01   \\
 8155$-$1901    &   1$-$38510    &   53.023305    &    0.479390    &     0.02206    &        9.58    &      -18.41    &        1.60    &        0.57    &         1.6    &        2.78   \\
8156$-$12704$^{c}$    &  1$-$205845    &   55.265549    &   -0.031234    &     0.03663    &        9.66    &      -18.79    &        5.15    &        0.57    &         1.0    &        1.29   \\
 8252$-$1902    &  1$-$138164    &  146.091838    &   47.459851    &     0.02590    &        9.40    &      -17.84    &        0.78    &        0.59    &         6.0    &        8.60   \\
 8258$-$1901    &  1$-$277213    &  166.620758    &   44.850902    &     0.02103    &        9.62    &      -18.57    &        1.28    &        0.52    &         0.9    &       29.61   \\
 8312$-$1901    &  1$-$210846    &  246.281956    &   39.435726    &     0.02962    &        9.68    &      -18.49    &        2.30    &        0.62    &         6.0    &        5.07   \\
 8320$-$3704    &  1$-$519742    &  206.612456    &   22.076742    &     0.02757    &        9.59    &      -18.82    &        2.68    &        0.60    &         1.3    &       13.12   \\
 8335$-$3703    &  1$-$252126    &  218.335552    &   40.550214    &     0.01816    &        9.41    &      -17.32    &        0.62    &        0.78    &         6.0    &       12.30   \\
 8338$-$9101    &  1$-$491095    &  171.966423    &   21.409205    &     0.02182    &        9.22    &      -17.96    &        3.22    &        0.38    &         0.9    &        4.41   \\
 8449$-$1902    &  1$-$488575    &  168.457073    &   22.810678    &     0.02237    &        9.52    &      -17.85    &        1.22    &        0.66    &         2.2    &        2.76   \\
 8483$-$1901    &   1$-$93551    &  246.311080    &   48.721331    &     0.02074    &        9.25    &      -17.25    &        0.61    &        0.64    &         6.0    &        3.42   \\
 8566$-$1901$^{c}$    &  1$-$338579    &  113.771912    &   41.974320    &    -0.00042    &        7.30    &      -12.13    &        0.00    &        0.86    &         6.0    &        0.69   \\
 8602$-$6101    &  1$-$211105    &  247.762350    &   39.623737    &     0.03242    &        9.60    &      -17.88    &        2.31    &        0.76    &         0.8    &        5.49   \\
 8613$-$1902    &  1$-$177127    &  255.290409    &   33.608349    &     0.03706    &        9.57    &      -18.97    &        1.40    &        0.44    &         1.3    &       33.92   \\
$^a$ 8655$-$1902    &   1$-$29809    &  358.468819    &   -0.098731    &     0.02196    &        9.55    &      -18.15    &        1.37    &        0.64    &         1.8    &        2.75   \\
$^a$ 8711$-$1901$^{c}$    &  1$-$379255    &  118.266565    &   52.743201    &     0.01846    &        9.48    &      -17.85    &        0.53    &        0.65    &         4.7    &        0.66   \\
 8711$-$1902    &  1$-$379008    &  118.761278    &   52.227437    &     0.02317    &        9.66    &      -18.34    &        1.27    &        0.66    &         2.0    &        9.34   \\
 8936$-$3704$^{c}$    &  1$-$152769    &  117.175199    &   30.171755    &     0.03587    &        9.64    &      -18.39    &        1.79    &        0.60    &         6.0    &        1.47   \\
 8937$-$1901    &  1$-$201307    &  116.997855    &   29.190689    &     0.02671    &        9.54    &      -18.21    &        1.62    &        0.57    &         3.3    &       15.74   \\
$^a$ 8942$-$6101$^{c}$    &  1$-$230177    &  124.897858    &   26.362666    &     0.01999    &        9.60    &      -18.25    &        2.55    &        0.65    &         1.8    &        1.61   \\
 8982$-$3703    &  1$-$458092    &  203.190094    &   26.580376    &     0.04701    &        9.46    &      -18.93    &        1.24    &        0.43    &         2.4    &       87.01   \\
 8983$-$1901$^{c}$    &  1$-$457905    &  203.795827    &   25.946970    &     0.02594    &        9.42    &      -17.75    &        1.27    &        0.64    &         1.9    &        1.20   \\
 8992$-$1902    &  1$-$149501    &  173.721052    &   51.091462    &     0.02625    &        9.55    &      -18.05    &        1.94    &        0.62    &         1.5    &        2.81   \\
 8999$-$6104    &  1$-$148597    &  165.161806    &   50.201391    &     0.02317    &        9.66    &      -18.65    &        1.03    &        0.56    &         6.0    &        8.69   \\
 9002$-$3702    &  1$-$373057    &  222.190591    &   31.687042    &     0.03222    &        9.52    &      -17.75    &        1.86    &        0.76    &         1.9    &        4.80   \\
9193$-$12703    &  1$-$109270    &   46.685858    &   -0.408800    &     0.02523    &        9.46    &      -18.28    &        1.46    &        0.51    &         0.9    &        9.02   \\
 9487$-$6102    &   1$-$45131    &  122.065499    &   45.284997    &     0.03979    &        9.55    &      -19.96    &        7.31    &        0.25    &         1.4    &       14.26   \\
 9505$-$3701$^{c}$    &  1$-$413676    &  139.139355    &   26.799501    &     0.02423    &        9.56    &      -17.99    &        2.23    &        0.62    &         2.1    &        1.68   \\
9509$-$12701    &  1$-$298640    &  123.265370    &   24.566993    &     0.02049    &        9.65    &      -18.52    &        3.53    &        0.55    &         2.1    &       13.75   \\
 9872$-$1902    &  1$-$322575    &  233.815569    &   41.570527    &     0.02827    &        9.66    &      -18.24    &        1.12    &        0.67    &         1.1    &        4.92   \\
10001$-$1901    &   1$-$55567    &  133.330028    &   57.041155    &     0.02575    &        9.45    &      -17.63    &        1.45    &        0.74    &         0.7    &       14.23   \\
\hline
 \multicolumn{11}{c}{The 16 off-nuclear AGN candidates}\\
\hline
 7960$-$9101    &  1$-$547621    &  259.206451    &   31.755753    &     0.02369    &        9.60    &      -18.86    &        3.54    &        0.47    &         1.3    &        7.68   \\
8317$-$12705   &  1$-$259104    &  193.855953    &   43.938064    &     0.03873    &        9.62    &      -19.24    &        5.65    &        0.43    &         1.4    &        5.21   \\
8335$-$12702    &  1$-$251764    &  215.487486    &   40.195802    &     0.01884    &        9.43    &      -18.53    &        3.96    &        0.39    &         1.6    &        4.49   \\
 8449$-$6104    &  1$-$488783    &  168.096446    &   23.169182    &     0.02291    &        9.55    &      -17.90    &        2.77    &        0.72    &         1.3    &        5.93   \\
 8464$-$6102    &  1$-$258752    &  186.000240    &   45.943987    &     0.02257    &        9.66    &      -19.02    &        2.84    &        0.43    &         0.8    &        6.36   \\
 8466$-$3702    &  1$-$277965    &  168.183562    &   45.187387    &     0.02336    &        9.45    &      -18.62    &        2.16    &        0.51    &         0.7    &        7.11   \\
8485$-$12705    &  1$-$209111    &  235.410126    &   48.274017    &     0.03796    &        9.67    &      -19.01    &        4.16    &        0.50    &         1.4    &        8.96   \\
8486$-$12704    &  1$-$209407    &  238.261763    &   46.767997    &     0.01951    &        9.39    &      -18.05    &        4.11    &        0.53    &         0.7    &        5.72   \\
 8623-9101    &  1$-$178698    &  311.347364    &    0.492529    &     0.01335    &        9.04    &      -18.03    &        2.19    &        0.47    &         1.6    &       10.05   \\
 8727$-$9102    &   1$-$51826    &   55.130622    &   -6.439925    &     0.02226    &        9.59    &      -19.08    &        3.29    &        0.35    &         0.7    &       24.46   \\
 8933$-$3704    &  1$-$457130    &  195.330499    &   27.860463    &     0.02741    &        9.43    &      -18.22    &        2.05    &        0.49    &         3.2    &        6.32   \\
8941$-$12701   &  1$-$164062    &  120.063604    &   27.589989    &     0.02295    &        9.20    &      -18.45    &        4.24    &        0.36    &         1.2    &        5.53   \\
 8987$-$3704   &  1$-$386264    &  136.294847    &   28.285333    &     0.02709    &        9.61    &      -18.31    &        3.22    &        0.68    &         1.8    &        7.15   \\
 8997$-$1901    &  1$-$149624    &  171.357399    &   52.622878    &     0.02710    &        9.56    &      -18.31    &        1.77    &        0.53    &         1.1    &        4.54   \\
 9000$-$1902    &  1$-$149695    &  172.886300    &   52.656030    &     0.03382    &        9.56    &      -18.18    &        0.88    &        0.61    &         2.6    &        7.65   \\
 9501$-$1902    &  1$-$385052    &  129.634703    &   24.898774    &     0.02595    &        9.52    &      -17.91    &        1.39    &        0.63    &         1.1    &        4.83   \\
\hline
 \multicolumn{11}{c}{The 11 AGN candidates  using the spectra in the central 1\arcsec.5 radius}\\
\hline
8335$-$12704    &  1$-$592984    &  215.718400    &   40.622597    &     0.01819    &        9.42    &      -17.76    &        3.86    &        0.54    &         1.2    &        3.98   \\
 8449$-$9101    &  1$-$488706    &  169.364330    &   23.346727    &     0.02232    &        9.27    &      -17.82    &        3.87    &        0.51    &         1.4    &        2.52   \\
 8601$-$6102    &  1$-$135512    &  247.711832    &   40.024799    &     0.02796    &        9.51    &      -17.89    &        3.44    &        0.57    &         2.8    &        3.59   \\
 8606$-$3704    &  1$-$136305    &  255.915543    &   36.384934    &     0.02467    &        9.52    &      -17.86    &        2.67    &        0.66    &         1.1    &        2.76   \\
 8612$-$9102    &   1$-$96154    &  254.397303    &   39.286089    &     0.03335    &        9.56    &      -18.34    &        2.92    &        0.57    &         1.0    &        4.12   \\
 8711$-$3703    &  1$-$379348    &  119.183464    &   52.989047    &     0.01784    &        9.05    &      -17.03    &        1.77    &        0.57    &         1.7    &        1.05   \\
 8941$-$9101    &  1$-$163975    &  120.203150    &   26.986272    &     0.02284    &        9.69    &      -18.35    &        3.50    &        0.64    &         3.8    &        0.88   \\
 8997$-$12703    &  1$-$149597    &  170.483361    &   52.850968    &     0.03370    &        9.63    &      -17.93    &        3.99    &        0.79    &         0.7    &        4.41   \\
 9486$-$3704    &   1$-$72128    &  122.052017    &   39.660395    &     0.01292    &        9.51    &      -17.94    &        1.23    &        0.64    &         1.1    &        1.24   \\
 9512$-$6103    &   1$-$53987    &  139.925621    &    1.325809    &     0.01650    &        9.37    &      -17.57    &        2.21    &        0.63    &         0.6    &        1.97   \\
 9868$-$12701    &  1$-$321290    &  218.468781    &   45.761149    &     0.03735    &        9.51    &      -18.34    &        3.53    &        0.85    &         1.5    &        3.39   \\
\hline
 \end{longtable}

\subsubsection{Control Sample}
To our purpose of exploring whether AGNs have affected the star formation of their host galaxies, we compare the stellar populations of AGN sample with those of control sample.  The control galaxies are selected from the remaining 837 dwarf galaxies (not including the 60 AGN candidates). The first criterion we adopt to select control galaxies is 
 \begin{center}
 \begin{equation}
 \Delta m= \left| {\rm log}\  M_{\star,\rm AGN}-{\rm log}\  M_{\star,\rm ctr} \right| \leqslant0.1,
 \end{equation}
 \end{center}
 where $M_{\star,\rm AGN}$ and $M_{\star,\rm ctr}$ represent the stellar mass of a member galaxy in the AGN and control samples, respectively. The reason why we use the stellar mass as the primary criterion is that the physical properties (such as star formation rate, stellar age, and metallicity) of nearby galaxies are strongly correlated with their stellar masses (e.g., \citealt{Kauffmann+etal+2004, Tremonti+etal+2004, Wuyts+etal+2011, Andrews+Martini+2013}).
 
 For each AGN, there might be more than one normal dwarf galaxies meeting the above mass criteria, and thus for those matched sources we further limit its redshift ($z$) to

  \begin{center}
 \begin{equation}
\Delta z= \left | \frac{z_{\rm AGN}- z_{\rm ctr}}{z_{\rm AGN}} \right| \leqslant 0.1.
 \end{equation}
 \end{center}
 We have then selected three control galaxies with the closest match in stellar mass for each AGN.  Different AGNs are allowed to share the same control galaxy.  For two AGN candidates (8982-3703 and 9486-3704) we have to adopt $\Delta z \leqslant$0.2 to select 3 control galaxies, and for one AGN candidate (8566-1901) we find no control galaxies.

We futher check if the galactic properties of the control sample are indeed compatible with those of the AGN sample.  As shown in  Figure \ref{comparefigure}, we show the distributions of  redshifts, stellar mass, absolute g band magnitude, effective radius, the $g-r$ color and  S\'{e}rsic  index for the two samples.  We can find that the distributions of the galactic properties are almost similar except  for the  $g-r$ color.  The median values of the $g-r$ color for the AGN and control samples are 0.58 and  0.44, respectively.    We will further discuss the effects of the different $g-r$ color in  \S\ 4.1.

\subsection{Spectra Stacking}
MaNGA provides spatially resolved spectra for every target galaxy, while  the spectra of outer regions have low S/N (especially for these dwarf galaxies), which could cause large uncertainty in deriving the SFHs. 
\cite{Cid+etal+2005} have found that the derived mean stellar ages ($\langle \log t_{\star}\rangle_{M}$ and $\langle \log t_{\star}\rangle_{L}$, see \S3) depend on the S/N of the input spectra with the  rms  less than 0.1 dex for ${\rm S/N} > 10$.

In order to increase the S/N, we have used elliptical annuli to radially bin the datacube of each galaxy, as shown in Figure \ref{figage} .  The width of the elliptical annuli is determined by the $Re$ (NSA\_ELPETRO\_TH50\_R) and the S/N (=mean/rms flux) measured in the relatively clean window between 4730 and 4780\r{A}. Each pixel size is 0$\arcsec$.5, therefore, the annular  width we set is equal to (0.5/$Re$+0.01) $Re$, which ensure that the minimum width is larger than 0$\arcsec$.5.   The width is increased until the binned spectrum has a S/N$\geq$10. We note that for some sources the central spaxel has a S/N$>$10, whereas the most outer region can not reach S/N$>$10.  During the binning process, we visually checked the contamination from foreground stars and/or neighbor galaxies, and masked those spaxels which could severely affect the binned spectra. However,  we do not remove those spaxels located within the most outer region since the binned spectra generally have  S/N$<$10 (see the green elliptical annuli in Figure \ref{figage}  ) and are not included in our analysis.

 \subsection{Stellar Population Synthesis}
 In order to derive the stellar populations of each binned spectrum for these dwarf galaxies, we employ the stellar population synthesis code, STARLIGHT (\citealt{Cid+etal+2005}, hereafter C05; \citealt{Mateus+etal+2006, Asari+etal+2007}).  This code searches the best match between  the observed spectra  and the model spectra by linearly  combining $N_{\star}$ simple stellar populations (SSPs), which were published by \cite{Bruzual+Charlot+2003} using the evolutionary synthesis models with the \cite{Salpeter+1995} initial mass function, Padova-1994 models, and the STELIB library (\citealt{LeBorgne+etal+2003}).   In this work, the model library comprises $N_{\star}=100$ SSPs which span 25 ages (from 1 Myr to 18 Gyr)  and 4 metallicities ($Z$=0.005, 0.02, 0.2, and 0.4$Z_{\odot}$), following \cite{Zhao+etal+2011} and \cite{Cai+etal+2020}. The SSPs are normalized at an arbitrary  $\lambda_0$ wavelength, reddened by a intrinsic extinction ($A_{V, \star}$) assuming a foreground screen dust geometry  with  \cite{Calzetti+etal+1994} extinction law where $R_{V}=4.05$ (\citealt{Calzetti+etal+2000})  and weighted by  the population vector $\bm{x}$ which represents the fractional contribution of each SSP to the total  synthetic flux at the normalized $\lambda_0$.

 Prior to the fitting,  the Galactic extinction law of \cite{Cardelli+etal+1989} and \cite{ODonnell+1994} with  $R_{V}=3.1$ is adopted to correct the observed spectra for  Galactic extinction with the $A_V$ values from \cite{Schlegel+etal+1998} as listed in the NASA/IPAC Extragalactic Database (NED).  Meanwhile, the observed spectra are corrected for redshift  and are rebinned to 1\AA\    sampling with Cubic Spline Interpolation.  The normalization wavelength we adopted  is  $\lambda_0$=4020\r{A} and the S/N of the observed spectra is calculated in the range free of emission lines between 4730 and 4780\AA.    The region of the spectra  used to fit is from 3800 to 8900\AA\  with masks of  20-30\AA\ around the obvious emission lines, and  
 the fitted regions are weighted equally, expect for the stellar absorption features (e.g., Ca\uppercase\expandafter{\romannumeral2}$K\lambda3934$; Ca\uppercase\expandafter{\romannumeral2} triplets), which are the strongest and less affected by nearby emission lines. More detailed informations about the synthesis process are presented in C05, \cite{Mateus+etal+2006} and \cite{Zhao+etal+2011}.

\begin{figure}[t!]
\centering
\includegraphics[width=0.85\textwidth]{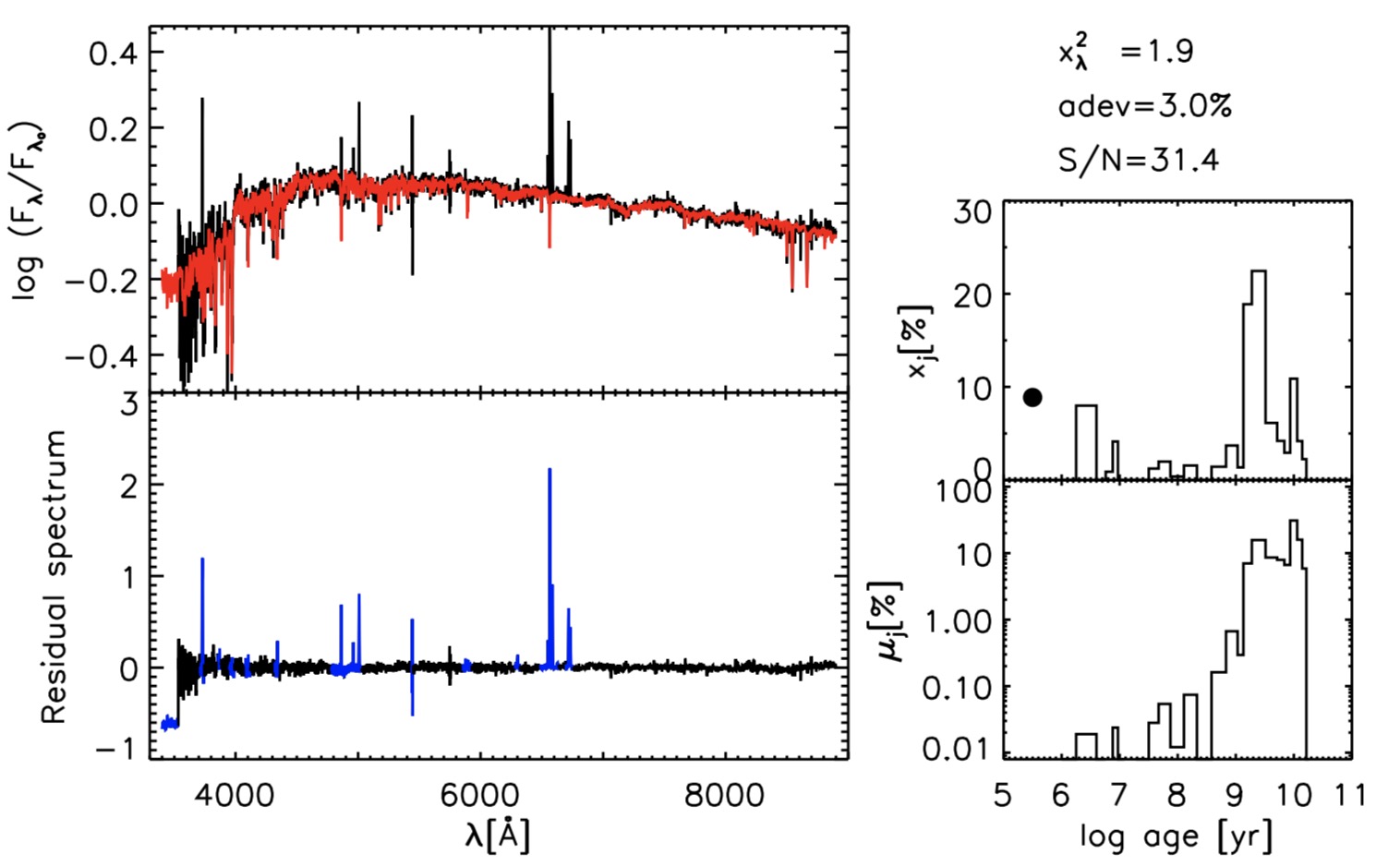}
\caption{ Results of the spectral fitting for the stacked spectrum (0.01$-$0.21 Re) of 9193-12703.   The logarithm of the  observed (black) and the synthetic (red) spectra are  shown in the top left panel  and the spectra are normalized to $\lambda_0\sim4020$\r{A}. The bottom left panel displays the residual spectrum  and the masked regions are replotted with a blue line. The right panels show the population vector binned in the 25 ages of SSPs used in the model library. The flux fraction vector and mass fraction vector are shown in  top  panel and bottom panel, respectively.  The power-law component $x_{\rm AGN}$ is plotted with an (arbitrary) age of $10^{5.5}$ yr and marked by a black point.} 
\label{figexample}
 \end{figure}

In addition, it is essential to add a non-stellar component (function of the form $F_{\upsilon} \propto \upsilon^{-1.5}$ )  during the fitting process to represent the contribution of an AGN featureless continuum  to the stellar base for the AGN sample, as same as previous works (e.g., C04; \citealt{Koski+1978, Riffel+etal+2009}).   As pointed out in C04, the  STARLIGHT code is incapable of discriminating between a $F_\nu \propto \nu^{-1.5}$ power-law continuum and the spectrum of a dusty starburst, which could cause an overestimation of the mean stellar age. We re-ran the code without adding an AGN featureless continuum for all binned spectra and compared the returned stellar ages with those with an  AGN featureless continuum.  The stellar ages  do not change much except for the binned spectra in five galaxies,  while this does not affect our conclusion.   We also investigated the fitted $A_V$  as in \cite{Cai+etal+2020} to avoid any unreasonable results.  We  re-ran the code by setting the lower limit of $A_V=0$ for about 100 binned spectra with their fitted $A_V<0$, among which about 40\% sources have their $A_V<-0.1$. The re-fitted results have no significant change when the $A_V$ is close to zero, and the re-fitted results are adopted  for these binned spectra.   In Figure~\ref{figexample}, we show a typical example of our fitting results for the stacked spectrum of 9193-12703.

\section{RESULTS and ANALYSIS}

According to the fractional contribution of flux and mass from different SSPs returned by STARLIGHT, we could calculate the  mean stellar age, mean stellar metallicity, which  are the most important stellar population parameters.  We calculate the light- and mass-weighted mean stellar ages using the equations described  in C05, i.e.
\begin{center}
 \begin{equation}
      \langle {\rm log}\   t_{\star}\rangle_{L}=\sum\limits_{j=1}^{N_{\star}} x_{j}\  {\rm log} \ t_{j},
 \end{equation}
 \end{center}
for the light-weighted mean stellar age, and
 \begin{center}
 \begin{equation}
      \langle {\rm log}\  t_{\star}\rangle_{M}=\sum\limits_{j=1}^{N_{\star}} \mu_{j}\ {\rm log}\  t_{j},
 \end{equation}
 \end{center}
for the mass-weighted mean stellar age. Here $t_{j}$  is the age value of the jth base, $N_{\star}$ is the total number of SSPs, and $x_j(\mu_j)$ represent the light (mass) fraction of the jth base.  As described in C05 and \cite{Zhao+etal+2011}, $\langle t_{\star}\rangle_{L}$ is associated with the recent SFH, while  $\langle t_{\star}\rangle_{M}$  mainly depend on  the mass assembly history.

Similarly, we calculate the light- and mass-weighted mean metallicity  using 
\begin{center}
 \begin{equation}
      \langle {Z_\star} \rangle_{L}=\sum\limits_{j=1}^{N_{\star}} x_{j} Z_{j},
 \end{equation}
 \end{center}
and
 \begin{center}
 \begin{equation}
      \langle {Z_\star} \rangle_{M}=\sum\limits_{j=1}^{N_{\star}} \mu_{j} Z_{j},
 \end{equation}
 \end{center}
 where $Z_{j}$ is the metallicity value of the jth base.  

\begin{figure}[t!]
\centering
\includegraphics[width=0.65\textwidth]{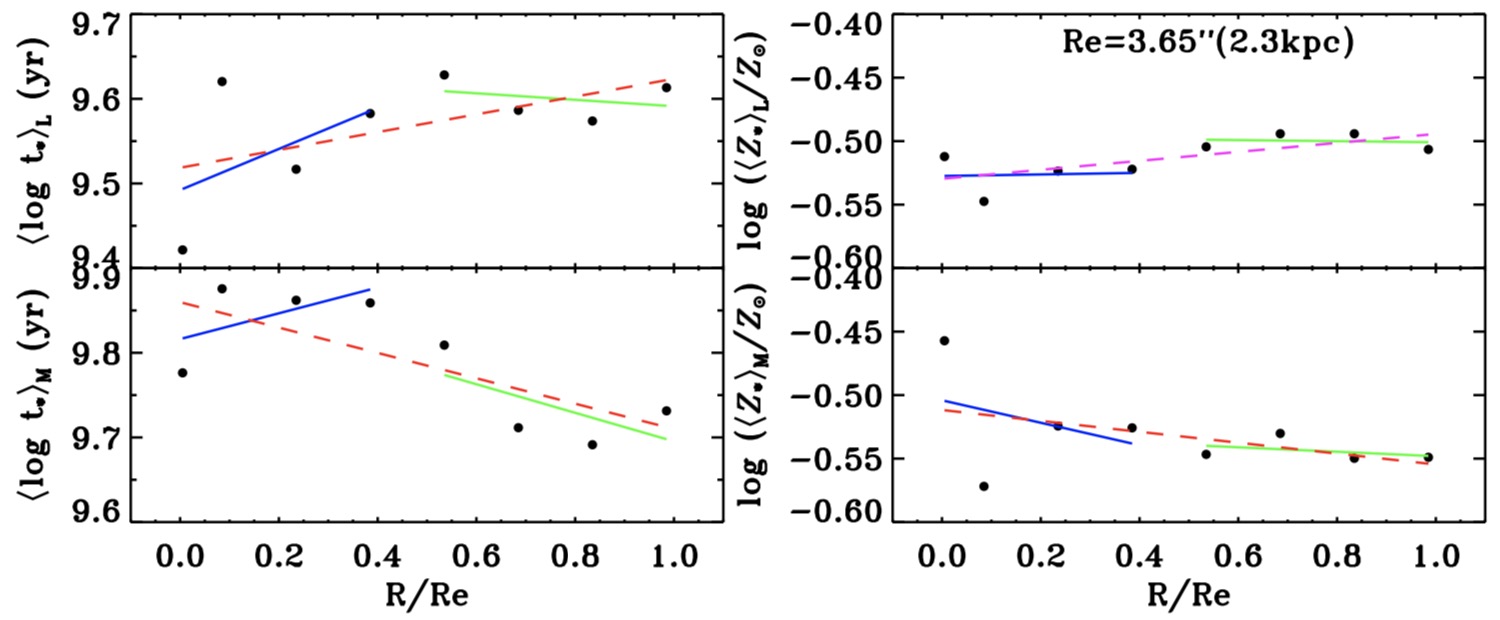}
\caption{A linear fitting example for MaNGA source 8312-1901.  The left two panels show  the linear fitting of the light-weighted mean stellar age (upper) and mass-weighted mean stellar age (bottom). The right two panels show the linear fitting of the light-weighted mean metallicity (upper) and mass-weighted mean metallicity (bottom). In each panel, The blue, green and red lines are the linear fit in $0-0.5 Re$,  $0.5-1 Re$ and $0-1 Re$, respectively.}
\label{gradientfigure}
 \end{figure}

\begin{figure*}[t!]
\centering
\includegraphics[width=0.85\textwidth]{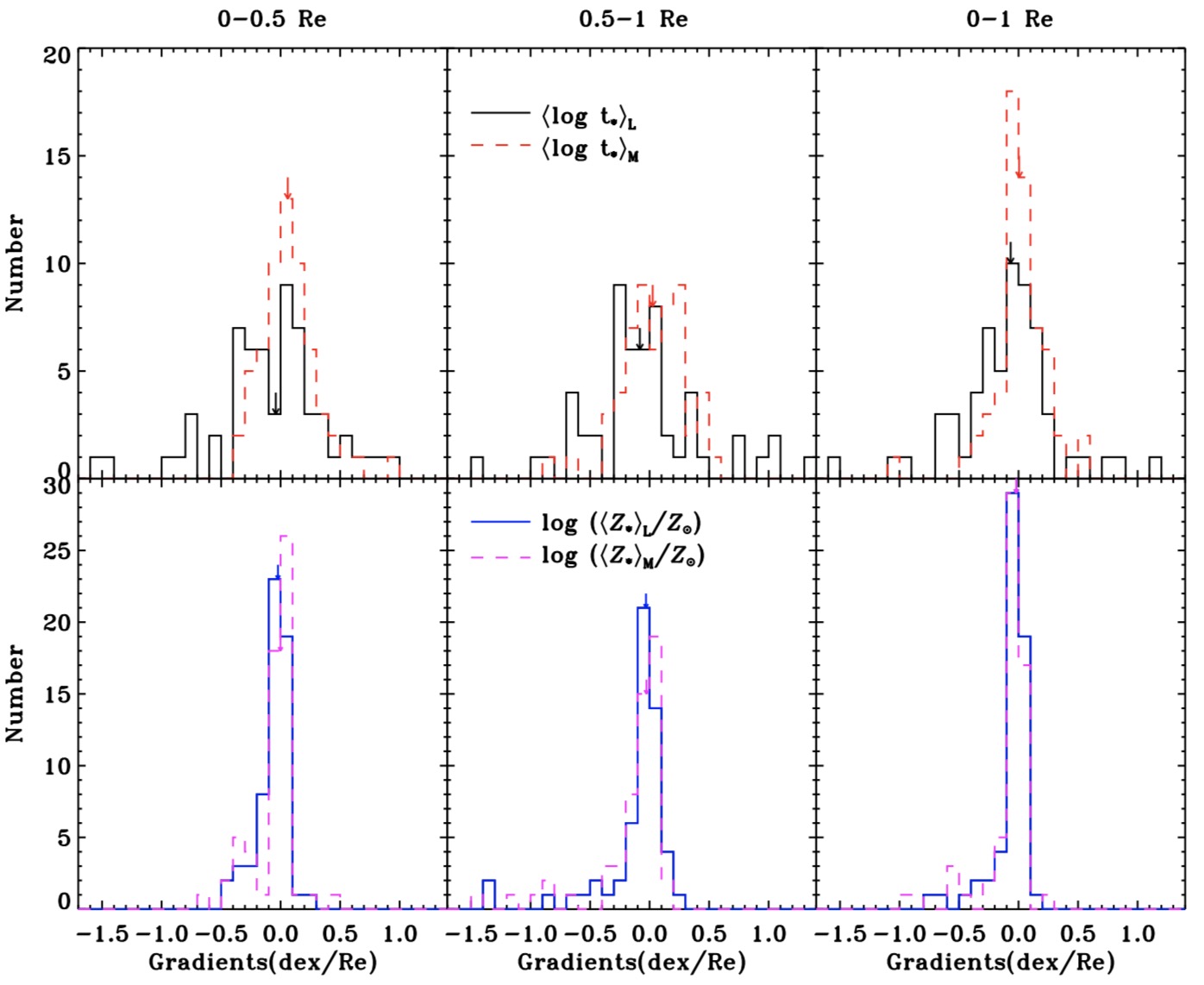}
\caption{ Number distributions of the gradients of the stellar age (upper) and stellar metallicity (bottom).   The left, middle and right panels show the distributions of the gradients in 0$-$0.5 $Re$, 0.5$-$1 $Re$ and 0$-$1 $Re$, respectively.   Arrows in each panel represent the median values of gradients.}
\label{gradient_dis}
\end{figure*}

According to the spatially resolved mean stellar age and mean stellar metallicity, we get the radial age  and metallicity profiles for each AGN-host dwarfs.    The low-level active galaxies should have effects on the stellar populations  in the nuclear regions.  The IFU of all MaNGA galaxies are covered to at least 1.5 $Re$, however, the radial profiles could not reach 1.5 Re due to poor S/N for the dwarf galaxies.  Some previous works (e.g., \citealt{ Mallmann+etal+2018}) have adopted 0.5 $Re$ as the division to make comparison  between  active galaxies and no-active galaxies. Therefore,   we measure the gradients of the profiles of individual galaxies in 0$-$0.5 $Re$, 0.5$-$1 $Re$, and 0$-$1 $Re$, respectively. We adopt a straight line to fit the data points with the following equation

\begin{center}
 \begin{equation}
    y=y_o+\triangledown\times x,
 \end{equation}
 \end{center}
where y=$\langle \log t_{\star} \rangle (\log (\langle Z_{\star}\rangle/Z_\odot))$,  the stellar age (metallicity), x=r/Re,  the normalized radius, and $\triangledown=\triangledown_{t}$ ($\triangledown_{Z}$), the gradient of stellar age (metallicity) in units of dex/Re.
 In Figure \ref{gradientfigure}, we take MaNGA source 8312-1901 as an example to show our fitting.

 Based on the fractional contribution of each SSP,  we can reconstruct the SFHs.  However, as discussed in C05, the  individual components of $\bm{x}$ returned by STARLIGHT have large uncertainties. Therefore, we adopt the same method with previous work (e.g., C05, \citealt{Zhao+etal+2011}, Cai et al. 2020), which binning vectors of $\bm{x}$  into three components, i.e. the young ($t< 10^8$yr), intermediate ($10^8 < t < 10^9$ yr), and old ($t > 10^9$ yr) populations. The  uncertainties of these three components are less than  0.05, 0.1, and 0.1, respectively, for ${\rm S/N} >10$. For each binned spectrum of the dwarf galaxies, the individual  components $\bm{x}$ are binned  into these three components ($x_{\rm Y}$, $x_{\rm I}$ and  $x_{\rm O}$), respectively.   In order to measure the gradients of the $x_{\rm Y}$($\triangledown_{x_{\rm Y}}$), we further adopt a straight line to  fit the $x_{\rm Y}$ for each AGN and control galaxy in  0$-$0.5 $Re$, 0.5$-$1 $Re$, and 0$-$1 $Re$.

\subsection{Radial stellar age}

In the upper panel of Figure \ref{gradient_dis}, we show the distributions of $\triangledown_{t}$ for the AGN sample.    The AGN sample has a wide and similar distribution in $\triangledown_{t_L}$($\triangledown_{t_M}$), ranging from $\sim$ -1.6(-0.5) to $\sim$ 1.0(1.0) in 0$-$0.5 $Re$,  $\sim$-1.5(-1.0) to $\sim$1.1(0.5) in 0.5$-$1 $Re$ and $\sim$-1.6(-1.0) to $\sim$1.2(0.6) in 0$-$1 $Re$.  
 The median values of $\triangledown_{t_L}$($\triangledown_{t_M}$) are   -0.04$\pm$0.36\footnote{Throughout the paper, the uncertainty of a median value is estimated using $1.48\times{\rm MAD}$ assuming normally distributed noise, where MAD is the median value of the absolute deviations from the median data.} (0.06$\pm$0.21) in 0$-$0.5 $Re$, -0.08$\pm$0.30 (0.03$\pm$0.26) in 0.5$-$1 $Re$ and -0.07$\pm$0.26 (0.00$\pm$0.14) in 0$-$1 $Re$, respectively.  The gradients of individual galaxies in different regions are different. However, the median values of the gradients are all close to 0 and this suggests that the overall behavior of the $\langle \log t_\star \rangle_L$ profiles for these AGN-host dwarfs are almost flat. We will further show the co-added radial stellar age in next section.

 Regrading the radial metallicity profiles,  the gradients of the individual source are also diverse. However,  the profiles tend to be flat or negative with few sources having the gradients greater than 0.2.  Meanwhile,  the median values of  $\triangledown_{Z_L}$($\triangledown_{Z_M}$)   are   -0.02$\pm$0.07 (0.00$\pm$0.04) in 0$-$0.5 $Re$,  -0.03$\pm$0.09 (-0.03$\pm$0.09) in 0.5$-$1 $Re$  and  -0.02$\pm$0.05 (-0.02$\pm$0.04) in 0$-$1 $Re$. Therefore, the metallicity profiles tend to be negative but close to zero. We will further discuss the  overall behavior of the stellar metallicity in next section.

 \begin{figure*}[ht!]
\centering
\includegraphics[width=0.9\linewidth]{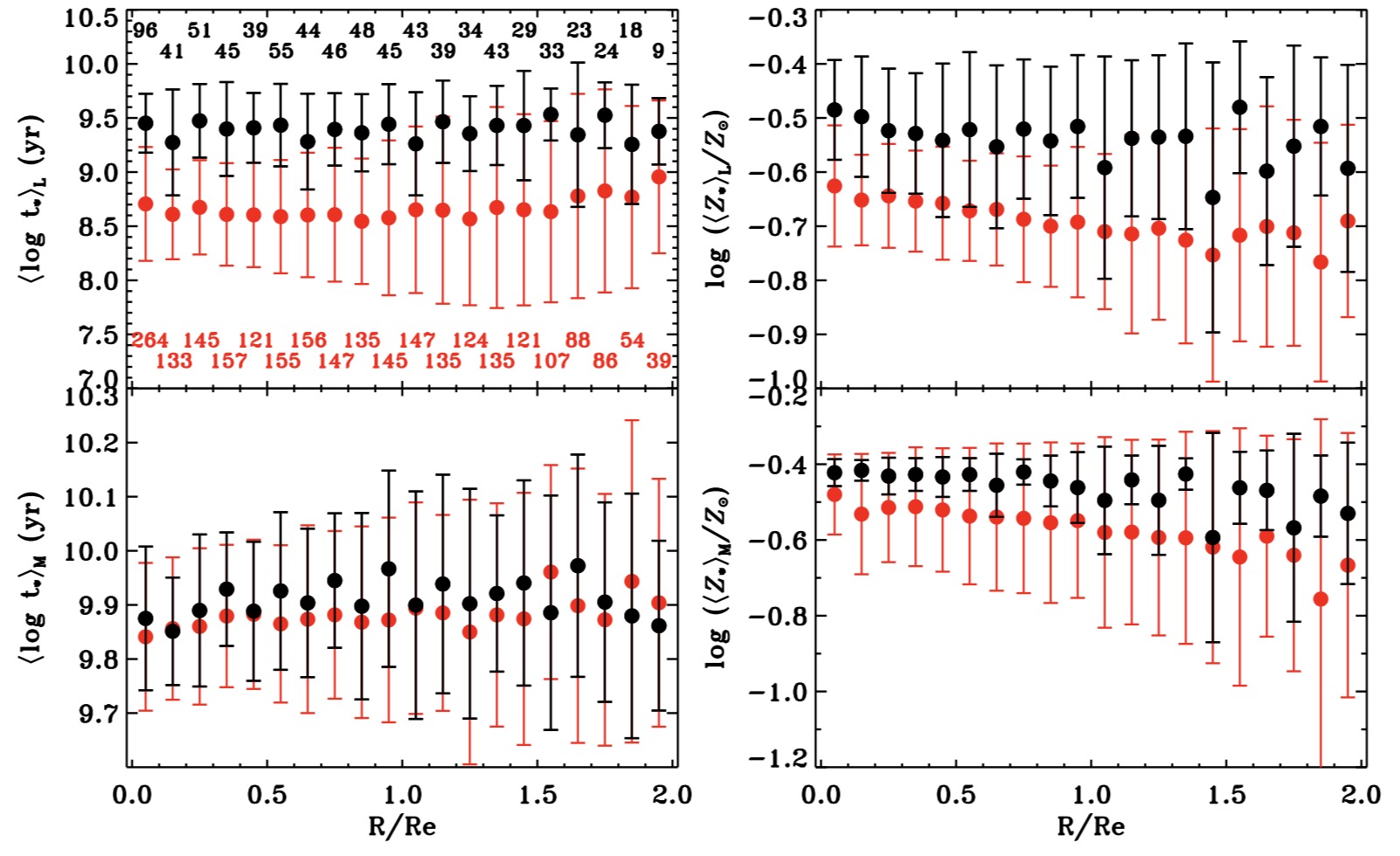}
\caption{ Co-added stellar age and metallicity profiles for the  AGN  and control samples. The left two panels show the co-added light- and mass-weighted mean stellar age profiles. The right two panels show co-added light- and mass-weighted metallicity profiles. The dots with error bars show the median value and 1$\sigma$ dispersion. The  co-added  radial profiles for the AGN and control samples are shown with black  and red colors, respectively.     In the upper left  panel, numbers of data points in each bin are labeled with black  and red colors for the AGN and control samples, respectively.   
}
\label{coaddfig}
\end{figure*}

 \begin{figure*}[ht!]
\centering
\includegraphics[width=0.9\linewidth]{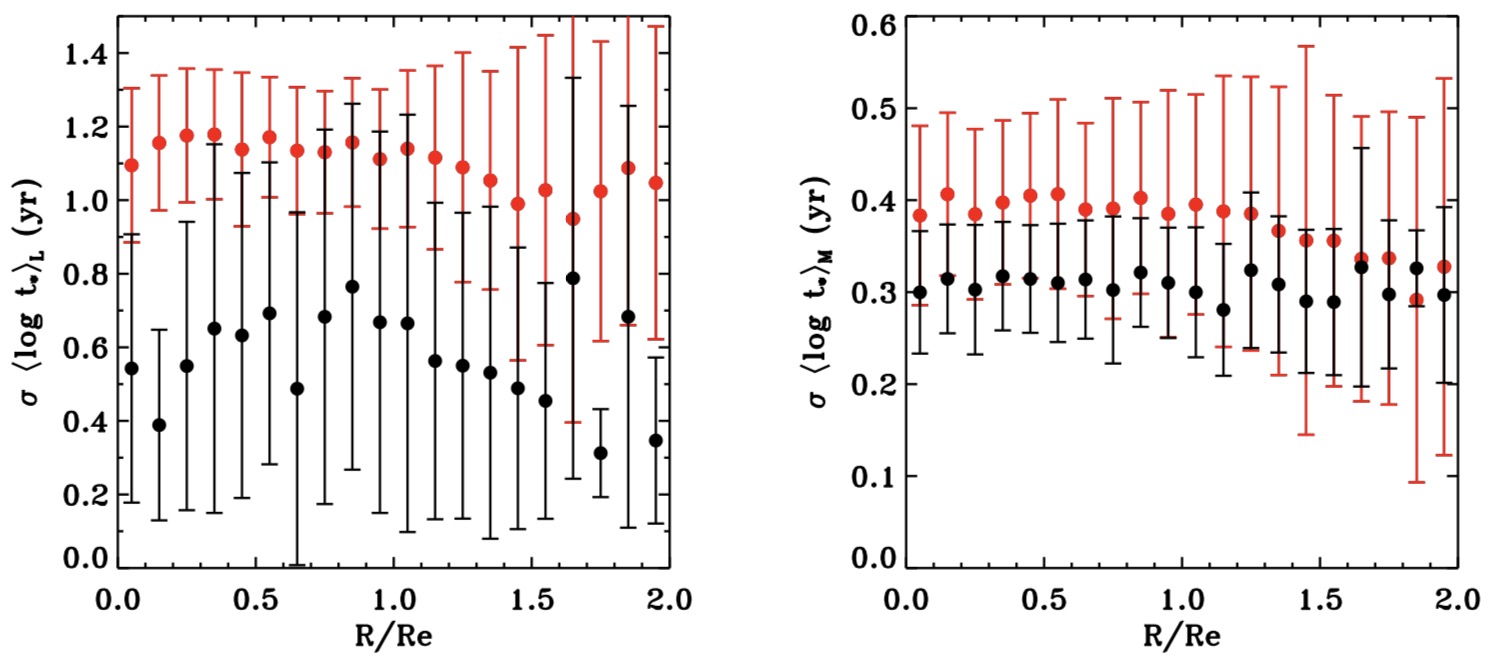}
\caption{ Co-added  standard deviations of the log age profiles for the  AGN  and control samples.  The dots with error bars show the median value and 1$\sigma$ dispersion. The  co-added  radial profiles for the AGN and control samples are shown with black and red colors, respectively.    
}
\label{coaddsig}
\end{figure*}

\subsection{Co-added radial properties}

In order to investigate AGN effects on host galaxies, we further compare the overall behavior of the stellar age and metallicity by plotting the co-added radial profiles for the  AGN  and control samples, as shown in Figure \ref{coaddfig}. Since we select 3 control galaxies for each AGN, and different AGNs are allowed to  share the same control galaxy.  Therefore we remove the repeated control galaxies and the control sample finally includes 147 individual dwarf galaxies. We radially bin the stellar age and metallicity with width of 0.1 effective radius ($Re$) for the two samples.   Due to the few data points in 1-2 $Re$ for the individual galaxy, we do not measure the gradients in 1-2 $Re$. However, the co-added radial profiles can reach the radius of $\sim$ 2 $Re$, which can be used to explore the properties in the outer region, and thus  we measure the gradients in the inner (0$-$1$Re$) and  outer (1$-$2 $Re$) regions.

  \begin{table}[tb]
\begin{center}
\caption{   The  gradients of the stellar age and stellar metallicity
\label{fittedresult}}
\begin{tabular}{ccccccc}
      \hline
      \hline   
         &\multicolumn{2}{c}{AGN} &\multicolumn{2}{c}{ AGN ( EW(H$\alpha) > 3$\r{A})} &\multicolumn{2}{c}{Control}  \\   
      \hline
         & inner  & outer  & inner & outer  & inner & outer  \\ 
         &  ( 0$-$1 Re) &  (1$-$2 Re) &( 0$-$1 Re)  & (1$-$2 Re) &( 0$-$1 Re)  & (1$-$2 Re)   \\ 
          \hline  
      $\langle \log t_\star \rangle_L$&-0.01&0.02&-0.05&-0.03&-0.12&0.31\\
      \hline
       $\langle \log t_ \star \rangle_M$ &0.08&-0.04&0.09&-0.03&0.03&0.05\\
       \hline
        $\log (\langle {Z_\star} \rangle_{L}/Z_\odot$)&-0.04&0.00&-0.03&0.03&-0.07&-0.01\\
      \hline
        $\log (\langle {Z_\star} \rangle_{M}/Z_\odot$) &-0.03&-0.06&-0.03&-0.05&-0.06&-0.13\\
       \hline
         \end{tabular}
      \end{center}
\end{table}

As shown in the bottom left panel of Figure  \ref{coaddfig}, the median values of  $\langle \log t_\star \rangle_M$ are both about 9.8$-$10.0 within 2 $Re$ for the two samples. Meanwhile, the 1 $\sigma$ dispersions of the $\langle \log t_\star \rangle_M$ shown by error bars range from $\sim$ 0.1 dex to $\sim$ 0.3 dex. Therefore, for these dwarf galaxies,  the stellar mass is mainly contributed by  the old population, which is similar to previous works (e.g., \citealt{Zhao+etal+2011, Cai+etal+2020}).   $\triangledown_{t_M}$ in the inner and  outer regions  are shown in Table \ref{fittedresult}. For both samples,  $\triangledown_{t_M}$ are close to 0 in the inner  and  outer regions, which suggest that  the $\langle \log t_\star \rangle_M$ profiles are generally flat. This is a natural result, as the present stellar mass is mainly contributed by old population for these objects.

Regarding $\langle \log t_\star \rangle_L$,     the gradients for the control sample are -0.12 and 0.31 in the inner and outer regions, respectively. The $\langle \log t_\star \rangle_L$ profile of the control sample show a ‘U’ shape curve with the minimum value located around 1$-$1.5 $Re$. A series of works (e.g., \citealt{Zheng+etal+2015, RuizLara+etal+2016, Zheng+etal+2017}) have found that the $\langle \log t_{\star}\rangle_{L}$ profiles of disk galaxies show a 'U' shape with the minimum value located around 1$-$1.5 $Re$. This may be caused by the stellar radial migration as suggested in numerical simulations(e.g., \citealt{Sellwood+Binney+2002, Roskar+etal+2008}).  However, the gradients for  the AGN sample are almost zero in the inner and  outer regions, as shown in Table \ref{fittedresult}.  One possible explanation might be that the AGN-host dwarf galaxies are almost quiescent due to  the AGN effects which resulting in an overall quenching of  the dwarf galaxies.

The median values of $\langle \log t_\star \rangle_L$ for the AGN sample are in the range of 9.3$-$9.5, which indicate that  these AGN-host dwarfs are almost quiescent in the last 2$-$3 ~Gyr within 2 $Re$.  The $\langle \log t_\star \rangle_L$ profile  for the AGN sample removing the 16 off-nuclear AGN candidates has no significant change.    For the control sample, the median values of $\langle \log t_\star \rangle_L$ range from $\sim$ 8.5 to $\sim$ 8.9 within 2 $Re$, which are significantly smaller than those for the AGN sample, with differences ranging from $\sim$ 0.4 dex to $\sim$ 0.9 dex.  
It can be found that the $\langle \log t_\star \rangle_L$ of the AGN sample  are very old and significantly older than the control sample selected with stellar mass as main criterion.  This indicates that these AGN-host dwarfs almost keep quiescent  within 2 $Re$. The reason for the weaker star formation in these AGN-host dwarf galaxies might be the lack of the gas.  
Some works (e.g., \citealt{Bradford+etal+2018, Sharma+etal+2020}) have found evidences of reduced H{\sc i} gas mass in isolated AGN-host dwarf galaxies, indicating a high degree of cold gas depletion.  The relatively older stellar ages of AGN host dwarfs imply that AGNs might  accelerate the evolution of galaxies by accelerating  the consumption of the cold gas, resulting in an overall quenching of  the dwarf galaxies.

However, some works (e.g., \citealt{StorchiBergmann+etal+2001, Rembold+etal+2017}) have found that the AGN activities in massive galaxies might be related to recent episodes of star formation in the nuclear region. \cite{Cai+etal+2020} also found not all AGN-host dwarf galaxies  are old.  Out of the 136 AGN-host dwarf galaxies, 15 galaxies (11\%) have $\langle \log t_{\star}\rangle_{L} < 8$ and 68 (50\%) galaxies have $8< \langle \log t_{\star}\rangle_{L} < 9$.  Further, a mild correlation is found between the SFHs and  \LOIII, for sources with $L_{\rm [O\,{\scriptsize \textsc{iii}}]}$$>$$10^{39}$~erg~s$^{-1}$.    An possible explanation is that these AGN-host dwarfs  stand in different stages of evolution.  At the early time when the host galaxies have abundant cold gas which can fuel the AGN and be used to star formation,  we  observe a young stellar age in at least the central region. The AGN-host dwarfs with old stellar age are in the late stages of evolution, in which the host galaxies almost have no recent star formation due to the lack of cold gas.  Meanwhile, as shown in \S\ 3.4, most of our AGN-host dwarf galaxies are  low-level AGNs, with only eight sources having $L_{\rm [O\,{\scriptsize \textsc{iii}}]}$$>$$10^{39.5}$~erg~s$^{-1}$.  The  weak strength of these AGNs might be also due to the lack of  gas, which is consistent with our previous explanation.

Therefore,  an entire picture might be that the stellar age of AGN-host dwarfs is young and the AGNs tend to be strong when the cold gas of host galaxies are abundant, which can feed black holes and be used for star formation.  As the fast consumption of cold gas by BHs or even AGN feedback, the dwarf galaxies gradually quench due to the lack of cold gas, and the AGNs also become weak.   Further, the quenched dwarf galaxies are likely the  low-mass analogues of the 'red geysers' suggested by (\citealt{Cheung+etal+2016}), which keep quiescent as the low-level  AGNs suppress the star formation by hampering the cooling of the gas (\citealt{Penny+etal+2018}).

The 1 $\sigma$ dispersion represents the diversity of the mean stellar age in the dwarf galaxies.   Since the $\langle \log t_\star \rangle_L$ of most AGN-host dwarfs are old  within 2 $Re$,  the 1 $\sigma$ dispersions of $\langle \log t_\star \rangle_L$  ($\sim$ 0.2$-$$\sim$ 0.6 dex) for the AGN sample are smaller than those ($\sim$ 0.4$-$$\sim$ 0.9 dex) for the control sample.  The 1 $\sigma$ dispersions of $\langle \log t_\star \rangle_L$  become larger with increasing $Re$,  especially for the control sample, which  might be caused by the less data points in high $Re$.    We further find that the 1 $\sigma$ dispersions of the control sample (0.4$-$0.9 dex) are larger than those of the low-mass sample ($\sim$ 0.5 dex) in \cite{Zheng+etal+2017}.  This is caused by the different methods used to calculate the stellar age.  As described in \S\ 3, this work use the  log $t_j$ before weighting, which may give more weight to younger stellar populations. For the control sample, we also calculate the $\langle \log t_\star \rangle_L$ by adopting the same definitions with \cite{Zheng+etal+2017}, and the 1 $\sigma$ dispersions of $\langle \log t_\star \rangle_L$ significantly decrease, ranging from  $\sim$ 0.2 dex to $\sim$ 0.5 dex.

In the right panel of  Figure  \ref{coaddfig}, we display the metallicity profiles for both samples. As shown in Table \ref{fittedresult}, $\triangledown_{Z_L}$  for the AGN (control) sample are -0.04 (-0.07) in the inner region and 0.0 (-0.01) in the outer region.  Meanwhile, $\triangledown_{Z_M}$ for the AGN (control) sample are -0.03 (-0.06) in the inner region and -0.06 (-0.13) in the outer region. The gradients of mean stellar metallicity  are close to 0 but slightly negative, which suggest that the metallicity profiles have a decreasing trend with increasing radius.   This is consistent with the results of other works (e.g., \citealt{Morelli+etal+2015, Zheng+etal+2017}), and the similar metallicity profiles for the AGN  and control samples  suggest that AGNs unlikely have a strong impact on the chemical evolution of the host galaxy, which corresponds with the results of \cite{Cai+etal+2020}.

According to C05, we can further investigate the SFH  by analyzing the light- and mass-weighted standard deviations of the log age, which are useful  to  identify whether galaxies  are dominated by a single population or  have bursty or continuous SFHs.  The definitions of these two parameters are
  \begin{center}
 \begin{equation}
      \sigma_{L}({\rm log}\  t_{\star})=\lbrack\sum\limits_{j=1}^{N_{\star}} x_{j}({\rm log} \ t_{j}-\langle {\rm log\ } t_{\star}\rangle_{L})^2\rbrack^{1/2}
 \end{equation}
 \end{center}
 and
  \begin{center}
 \begin{equation}
      \sigma_{M}({\rm log}\  t_{\star})=\lbrack\sum\limits_{j=1}^{N_{\star}} \mu_{j}({\rm log}\  t_{j}-\langle {\rm log}\  t_{\star}\rangle_{M})^2\rbrack^{1/2}.
 \end{equation}
 \end{center}

In Figure \ref{coaddsig}, we plot the radial median values of $\sigma_{L}$(log $t_{\star})$ and $\sigma_{M}$(log $t_{\star})$ for the AGN and control samples.  It can be found that these dwarf galaxies are not dominated by a single population, as the median values of the  $\sigma_{L}$({\rm log} $t_{\star})$ and $\sigma_{M}$(log $t_{\star})$ within 2 $Re$ are  greater than zero.  The median values of $\sigma_{L}$(log $t_{\star})$ tend to be larger than  $\sigma_{M}$(log $t_{\star})$. This indicates that the mass assembly history of these dwarf galaxies is relatively narrow, while  repeated/continuous star formation activities might exist during the lifetime of dwarf galaxies.   Further, there exist  differences in the median values of $\sigma_{L}$(log $t_{\star})$ between the AGN and control sample:  the median values of  $\sigma_{L}$(log $t_{\star})$ within 2 $Re$ range from 0.3 (1.0) to 0.8 (1.2) for the AGN (control) sample. This is a reasonable result, as the AGN sample is dominated by  old stellar populations.

\subsection{Dependence on the Host Galaxy Morphology}
\begin{figure*}[ht!]
\centering
\includegraphics[width=0.85\textwidth]{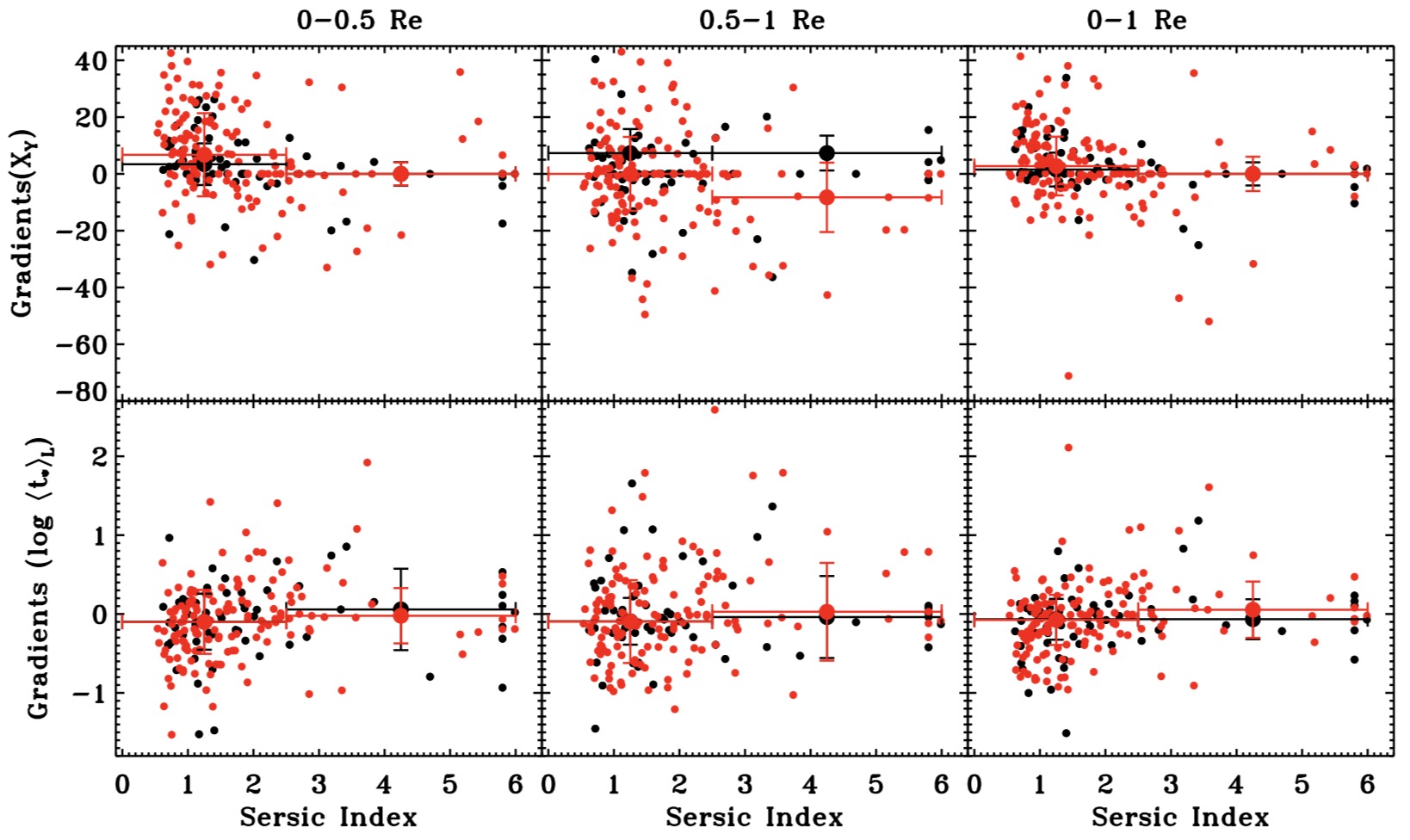}
\caption{Plot of S\'{e}rsic index $n$ of the dwarf galaxies versus $\triangledown_{x_{\rm Y}}$ (upper) and $\triangledown_{t_L}$ (bottom)   in 0$-$0.5 $Re$, 0.5$-$1 $Re$ and 0$-$1 $Re$, respectively.
In each panel, the AGN  and control samples are represented with the black and red circles, respectively.
 The bigger black (red) solid circles represent the median values of the gradients for early-type ($n>2.5)$ and late-type ($n<2.5$) AGN-host dwarf (control) galaxies,  with error bars in the $y$-axis giving the dispersion of the data within each bin.}
\label{indexage}
\end{figure*}

  The SFHs of galaxies are related to the galaxy morphology. A series of works have studied the relationship between the nuclear stellar population and the AGN host galaxy morphology. Most of works found no correlation between the nuclear star formation and the  host galaxy morphology (e.g., C04;\citealt{Cai+etal+2020}).   \cite{ Mallmann+etal+2018} further tried to study the spatially resolved stellar population of AGN host galaxies and found no significant differences between early- and late-type host galaxies.  In this paper, for these AGN-host dwarf galaxies,  we try to search the correlation between the gradients of SFHs and the  host galaxy morphology.

Since the detailed Hubble morphology of dwarf galaxies are not easy to recognized, we use the S\'{e}rsic index as galaxy morphology parameter. By adopting $n=2.5$ as the dividing line (e.g., \citealt{Barden+etal+2005}), the dwarf galaxies were classified as  early- ($n>2.5$) and late-type ($n<2.5$) galaxies.  As shown in Figure \ref{indexage}, we plot the S\'{e}rsic index $n$ against  $\triangledown_{t_L}$ and  $\triangledown_{x_{\rm Y}}$ with the median values of gradients overlaid. From the figure we can find that, for both the AGN  and control samples, the $\triangledown_{t_L}$ and  $\triangledown_{x_{\rm Y}}$  in 0$-$0.5 $Re$, 0.5$-$1 $Re$ and 0$-$1 $Re$ almost have no  correlation with $n$. Furthermore, we can find that the median values of $\triangledown_{t_L}$ and  $\triangledown_{x_{\rm Y}}$  almost have no differences between the AGN sample and the control sample except for the $\triangledown_{x_{\rm Y}}$  in 0.5$-$1 $Re$. However, the median values of of the $\triangledown_{x_{\rm Y}}$  in 0.5$-$1 $Re$ for both samples are  similar when we consider the errors. Therefore, for both samples, no obvious correlations between the gradients of SFHs within 1 $Re$ and the morphology are found.

\subsection{ \OIIII\ Luminosity of the AGN sample }
\begin{figure}[t!]
\centering
\includegraphics[ width=0.85\textwidth]{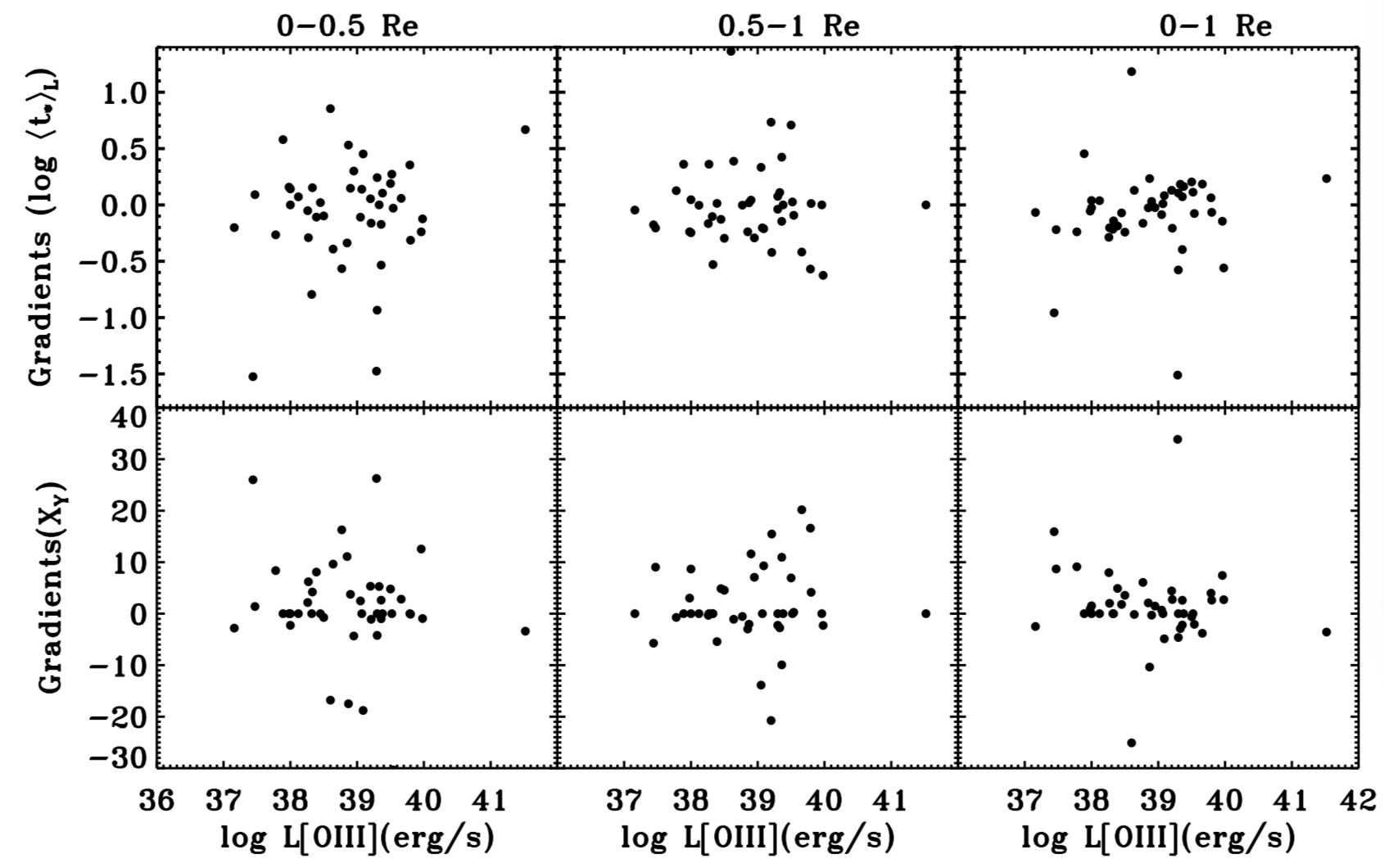}
\caption{Plot of the \OIII\ luminosity versus  $\triangledown_{t_L}$ (upper) and $\triangledown_{x_{\rm Y}}$ (bottom)  in 0$-$0.5 $Re$, 0.5$-$1 $Re$ and 0$-$1 $Re$, respectively. 
}
\label{AGNage12}
\end{figure}

As discussed in \S\  3.2, most of  AGN candidates keep quiescent due to the lack of gas, which suggest that  the  strength  of  these AGNs should be weak. Here we research the \OIIII\ luminosity of these AGN candidates, as \cite{Kauffmann+etal+2003} have shown that the \OIII\ line luminosity, \LOIII, is a good tracer of AGN activity (see also \citealt{Trump+etal+2015}).  We derive the  \OIIII\ luminosity for these AGN candidates by using the central 1\arcsec.5 radius spectra, and the \LOIII\  are corrected for extinctions, which are measured  from the observed H$\alpha$/H$\beta$ Balmer decrement (see section 3.7 of \citealt{Cai+etal+2020}).

As shown in Figure \ref{AGNage12},  out of 44 AGN candidates classified as AGNs using the spectra from the central 1\arcsec.5 radius , the \LOIII\  of 36 objects is in the range of $10^{36.5}-10^{39.5}$~erg~s$^{-1}$ (i.e. $10^3-10^{6}\,L_\odot$), and only one  object can be characterized as  ``strong AGN", as its  $L_{\rm [O\,{\textsc {\scriptsize {iii}}}]} \geq 10^7\,L_\odot$ (e.g., \citealt{Kauffmann+etal+2003}). This confirms that the AGN candidates are the low-level AGNs.  \cite{Cai+etal+2020} have found a mild (anti-) correlation between $x_{\rm Y}$ ($\langle \log t_{\star}\rangle_{L}$) and \LOIII, for sources with $L_{\rm [O\,{\scriptsize \textsc{iii}}]}$$>$$10^{39}$~erg~s$^{-1}$. Based on their work, the stellar populations of nuclear regions tend to be younger when AGN becomes stronger. This indicates that  a positive (negative) correlation might be found between  $\triangledown_{t_L}$( $\triangledown_{x_{\rm Y}}$) in the inner region and \LOIII.  As shown in Figure \ref{AGNage12}, no obvious correlation is found between  $\triangledown_{t_L}$( $\triangledown_{x_{\rm Y}}$) and \LOIII.  The Spearman correlation coefficients ($\rho$) are 0.09(-0.15), -0.02(0.15) and 0.23(-0.20) in 0$-$0.5 $Re$, 0.5$-$1 $Re$ and 0$-$1 $Re$, respectively, with $p$-values of $5.8\times10^{-1}$($3.3\times10^{-1}$), $8.9\times10^{-1}$($3.2\times10^{-1}$) and $1.2\times10^{-1}$($1.9\times10^{-1}$).  Most of the AGN candidates are low-level AGNs, therefore, no correlation between the  gradients and  \LOIII\  might be due to few strong sources.

\section{DISCUSSION}
\subsection{The selection effect of the AGN sample}
  \begin{figure}[t!]
 \centering
\includegraphics[width=0.47\textwidth]{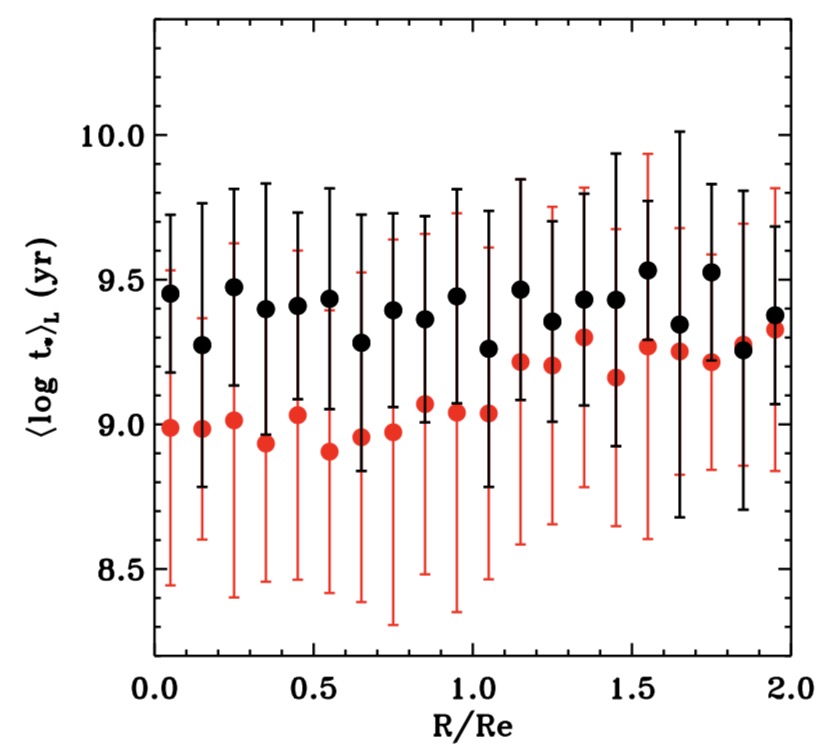}
\caption{ Co-added stellar age profiles for the  AGN  and new color-selected control samples. The dots with error bars show the median value and 1$\sigma$ dispersion. The  co-added  radial profiles for the AGN and control samples are shown with black  and red color, respectively.    }
\label{izcomfi}
\end{figure}

As shown in Figure \ref{comparefigure}, the median values of the $g-r$ color for the AGN and control samples are 0.58 and  0.44, respectively.  The redder color of this AGN sample is likely caused by a selection effect, which might miss the young AGN dwarf galaxies by BPT diagram due to the AGN radiation being severely contaminated by the strong radiation from H{\sc ii} regions. Therefore, it is essential to consider the effect of the different $g-r$ color when we compare the stellar ages between the AGN and control samples.

Fortunately, \cite{Zhao+etal+2011} have suggested that  the fitted $\langle \log t_\star \rangle_L$ has a good correlation with the $g-r$ color.   Based on their work (see Figure 9 in \citealt{Zhao+etal+2011}), the AGN sample is  0.14  mag (median) redder than the control sample in $g-r$ color, which causes that the  $\langle \log t_\star \rangle_L$ of the AGN sample is about 0.3$-$0.4 dex older than that of control sample.  However, as shown in Figure \ref{coaddfig}, the differences between the two samples in $\langle \log t_\star \rangle_L$ range from $\sim$ 0.4 dex to $\sim$ 0.9 dex within the 2 $Re$. The AGN sample is still $\sim$0.3$-$$\sim$0.6 dex older than the control sample within at least 1 $Re$ when we remove the effects of the $g-r$ color.

We further select a new control  sample adopting the stellar
mass  and $g-r$ color in the whole galaxy as our primary selection criteria.  The color criterium is
 
   \begin{center}
 \begin{equation}
   \left|(g-r)_{\rm AGN}-(g-r)_{\rm ctr}\right|\leqslant 0.05,
 \end{equation}
 \end{center}
where $(g-r)_{\rm AGN}$ and $(g-r)_{\rm ctr}$ represent the $g-r$ colors in the AGN and control samples, respectively.  For each AGN, we select the control galaxy which meeting the mass and color criteria.
Similar with the situation in  \S\  2.1.2, for each AGN, there might be more than one normal dwarf galaxy meeting the above criteria, and thus for those matched sources we further limit its redshift based on equation (3). Finally, we  choose the one with the closest match in $g-r$ color if there are still more than one matched objects.  

 \begin{figure}[t!]
\centering
\includegraphics[ width=0.47\textwidth]{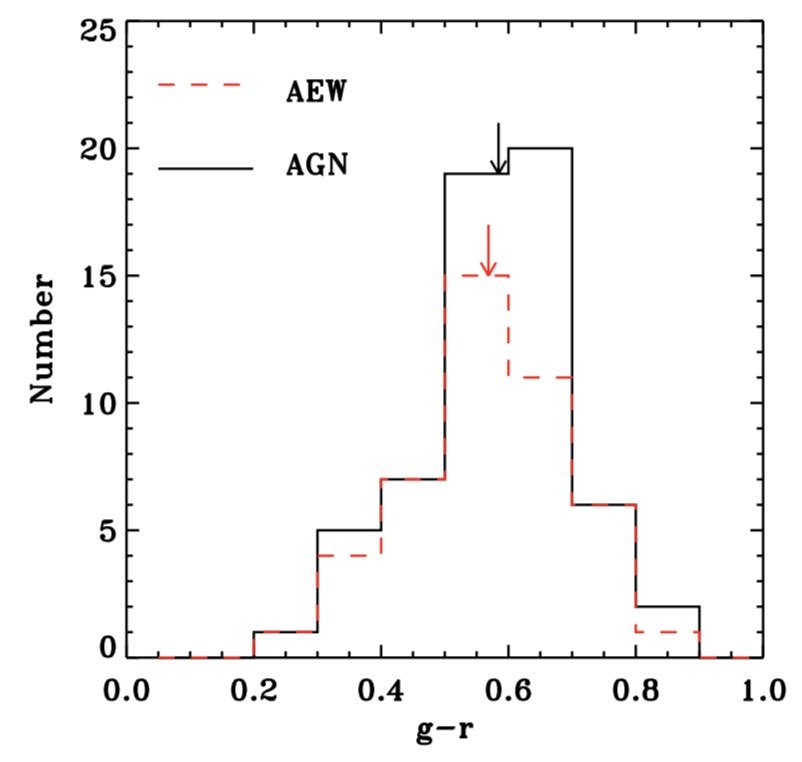}
\caption{ Number  dsistributions of g-r color for the AGN sample and the AGN sample with EW (H$\alpha) > 3$\r{A}.   The black solid line and red dashed line represent the AGN sample and  the AGN sample with EW (H$\alpha) > 3$\r{A}, respectively.   Arrows  represent the median values of the g-r color. }
\label{izcomfi2}
\end{figure}

In Figure \ref{izcomfi}, we show  $\langle \log t_\star \rangle_L$   profiles for the AGN and new color-selected control samples, respectively. The median values of  $\langle \log t_\star \rangle_L$  for the new control sample are about  9.0 in 0$-$1 $Re$ and larger than 9.0 in 1$-$2 $Re$.  The stellar ages of new color-selected control sample are   younger than AGN sample, especially in the 0$-$1 $Re$, with differences ranging  from $\sim$ 0.3 dex to $\sim$ 0.5 dex.   In the outer region, the differences of the stellar age tend to be small.   Therefore,  for these BPT-selected AGN-host dwarf galaxies, although the differences of the stellar age between the AGN and control samples become smaller when we consider the selection effect,  the stellar ages within at least 1 $Re$ are still $\sim$ 0.3 dex to $\sim$ 0.5 dex older than those of control galaxies.  This is agreement with our previous conclusion that  AGN might  cause the old stellar age in dwarf galaxy by  accelerating the consumption of the cold gas, but limited to the inner region.

\subsection{The  H$\alpha$ equivalent width of the AGN sample }

  The AGN sample is selected according to the BPT diagram, which might cause the selected AGN dwarf galaxies  are not  true AGNs. A series of works (e.g., \citealt{Binette+etal+1994, Sarzi+etal+2010, Cid+etal+2011,Papaderos+etal+2013, Singh+etal+2013}) have argued that the low-mass hot evolved stars might be the emission sources of the harder ionization field, which could be mistaken for AGN. The criteria (H$\alpha$ equivalent width, EW(H$\alpha) > 3$\r{A}) suggested by \cite{Cid+etal+2011}  are used to check if the selected AGN candidates are true AGNs.  For the 49 AGN candidates identified by the spatially resolved BPT diagram,  we re-construct the spatially resolved BPT diagram for those spaxels with $\rm S/N > 3$ and   EW(H$\alpha) > 3$\r{A} and   find that 40 sources are true AGNs. Further,  for  the 11 AGNs added by using the central 1\arcsec.5 radius spectra, there exist 5 galaxies  having EW(H$\alpha) > 3$\r{A}, as shown in  Table  \ref{agnsm}.  

 The $g-r$ color distribution of these 45 AGN candidates meeting EW(H$\alpha) > 3$\r{A} is shown in Figure \ref{izcomfi2}. The median value of the $g-r$ color is 0.57, which is still  redder than the control sample (0.44).  Therefore, the red $g-r$ color of  AGN sample is not caused by mistaking the low-mass hot evolved stars in the center as the  AGN. The selected AGN dwarf galaxies according to BPT diagram are indeed red galaxies. We further checked the co-added radial properties for these 45 AGN candidates and found that the stellar age and metallicity profiles  have no significant change,  as shown inTable  \ref{fittedresult}.

\section{summary}

We search AGNs using BPT diagram among a  parent sample including about 900  dwarf galaxies from MaNGA with stellar mass less than $5\times10^{9}$$M_{\odot}$.  We derive the spatial resolved stellar populations with STARLIGHT code and construct the radial stellar age/metallicity  profiles  for this AGN sample. In order to investigate AGN effects on dwarfs,
we further compare the co-added radial properties of the AGN sample with  those of a control sample. Our main results are summarized as follows:

\begin{enumerate}
    
    \item The median values of the co-added  light-weighted mean stellar age for the AGN sample are as old as 2$-$3 ~Gyr within 2 $Re$, while for the control sample, the median values of the co-added  light-weighted mean stellar age range from  $3 \times 10^8$ to $10^9 $yr, which are about 3 $-$7 times younger than the AGN sample. This indicates that these AGN-host dwarfs are almost quiescent due to the lack of gas.    
Further, most of our AGN candidates are low-level AGNs, as only one objects have its  $L_{\rm [O\,{\textsc {\scriptsize {iii}}}]} \geq 10^7\,L_\odot$. This is consistent  with the  old stellar age of the AGN-host dwarf galaxies, as the AGNs also become weak due to the lack of gas.

    \item  The co-added mass-weighted mean stellar age profiles for the AGN and control samples are both almost flat within 2 $Re$. Meanwhile, the median values of the co-added  mass-weighted mean stellar age for the AGN and control samples are  both about 10 Gyr within 2 $Re$, which suggests the stellar masses of  these dwarf galaxies  are mainly contributed by the old populations.  For the control sample, the co-added light-weighted mean stellar age profile shows a ‘U’ shape curve with a negative gradient (-0.12) in the inner region and a positive gradient (0.31) in the outer region. However, the co-added  light-weighted mean stellar age profile for the AGN sample is  almost flat within 2 $Re$.

    \item   For both the AGN and control samples, the $\triangledown_{t_L}$ and  $\triangledown_{x_{\rm Y}}$ in 0$-$0.5 $Re$, 0.5$-$1 $Re$ and 0$-$1 $Re$ almost have no correlation with the galaxy morphology.

     \item The co-added light-weighted standard deviation of the log age with 2 $Re$ are in the range of 0.3$-$0.8 (1.0$-$1.2) for the AGN (control) sample, while the co-added mass-weighted standard deviation of the log age have  a much narrower range (0.2$-$0.4) for both sample,  suggesting that these dwarf galaxies experience a continuous star-forming activities, but the stellar mass is assembled within a short period of time.

     \item The co-added mean stellar metallicity profiles have a decreasing trend with increasing radius for both sample, and the similar mean stellar metallicity profiles for both samples indicate that AGNs have no strong impact on the chemical evolution of the host galaxy.
  \end{enumerate}   

\normalem
\begin{acknowledgements}
We thank the anonymous referee for a careful reading and thoughtful comments which improved the paper. We thank Dr. Zhang Hong-Xin for a thorough reading of the paper and the useful suggestions which have improved the paper. The work is supported by the National Key R\&D Program of China grant No. 2017YFA0402704, the Natural Science Foundation of China (NSFC; grant Nos. 11991051, 11421303 and 11973039), and the CAS Pioneer Hundred Talents Program. The STARLIGHT project is supported by the Brazilian agencies CNPq, CAPES, and FAPESP and by the France–Brazil CAPES/Cofecub program. All the authors acknowledge the work of the Sloan Digital Sky Survey (SDSS) team. Funding for SDSS- IV has been provided by the Alfred P. Sloan Foundation and Participating Institutions. Additional funding towards SDSS-IV has been provided by the US Department of Energy Office of Science. SDSS-IV acknowledges support and resources from the Centre for High-Performance Computing at the University of Utah. The SDSS web site is www.sdss.org.
SDSS-IV is managed by the Astrophysical Research Consortium for the Participating Institutions of the SDSS Collaboration including the Brazilian Participation Group, the Carnegie Institution for Science, Carnegie Mellon University, the Chilean Participation Group, the French Participation Group, Harvard–Smithsonian Center for Astrophysics, Instituto de Astrofsica de Canarias, The Johns Hopkins University, Kavli Institute for the Physics and Mathematics of the Universe (IPMU)/University of Tokyo, Lawrence Berkeley National Laboratory, Leibniz Institut f{\"u}r Astrophysik Potsdam (AIP), Max-Planck-Institut f{\"u}r  Astronomie (MPIA Heidelberg), Max-Planck-Institut f{\"u}r Astrophysik (MPA Garching), Max-Planck-Institut f{\"u}r Extraterrestrische Physik (MPE), National Astronomical Observatory of China, New Mexico State University, New York University, University of Notre Dame, Observ{\'a}trio Nacional/MCTI, The Ohio State University, Pennsylvania State University, Shanghai Astronomical Observatory, United Kingdom Participation Group, Universidad Nacional Aut{\'o}noma de M{\'e}xico, University of Arizona, University of Colorado Boulder, University of Oxford, University of Portsmouth, University of Utah, University of Virginia, University of Washington, University of Wisconsin, Vanderbilt University and Yale University.
\end{acknowledgements}

\bibliographystyle{raa}
\bibliography{bibtex}

\end{document}